\documentclass[useAMS,usenatbib,usegraphicx]{mn2e}

\title[The Circumnuclear Region of M100]{The Star Formation History and Evolution of the Circumnuclear Region of M100}
\author[E. L. Allard et al.]{E. L. Allard$^{1}$\thanks{E-mail:
allard@star.herts.ac.uk}, J. H. Knapen$^{1}$, R. F. Peletier$^{2}$ and M. Sarzi$^{1}$\\
$^{1}$Centre for Astrophysics Research, University of Hertfordshire, College Lane, Hatfield, Herts, AL10 9AB, UK\\
$^{2}$Kapteyn Astronomical Institute, University of Groningen, Postbus
  800, 9700 AV, Groningen, The Netherlands}

\begin{document}

\date{Accepted 1988 December 15. Received 1988 December 14; in original form 1988 October 11}

\pagerange{\pageref{firstpage}--\pageref{lastpage}} \pubyear{2002}

\maketitle
\label{firstpage}
\begin{abstract}
\emph{Context}
Star-forming nuclear rings in barred galaxies are common
in nearby spirals, and their detailed study can lead to important
insights into the evolution of galaxies, their bars, and
their central regions. We present integral field spectroscopic observations obtained with
SAURON of the bar and circumnuclear region of the barred spiral galaxy
M100, complemented by new {\it Spitzer Space Telescope} imaging of the region.\\
\emph{Aims}
We use these data to enhance our understanding of the formation,
evolution, and current properties of the bar and ring.\\
\emph{Methods} 
We derive the kinematics of the gas
and the stars and quantify circular and non-circular motions using
kinemetry. We analyse this in conjunction with the optical and
infrared morphology, and our previously published dynamical
modelling. By comparing line indices to simple stellar
population models we estimate the ages and metallicities of the
stellar populations present within the region, especially in and
around the ring.\\
\emph{Results} 
The stellar and gaseous velocity fields are remarkably
similar, and we confirm that the velocity fields show strong evidence
for non-circular motions due to the bar and the associated density wave. These
are strongest just outside the nuclear ring, where our kinemetric
analysis indicates inflow across the spiral armlets and into the ring
region.  The line strength maps all indicate the presence of 
a younger population within
this ring, but detailed modelling of the line strengths shows that in
addition to this young population, old stars are present. These old
stars must have been formed in an event of massive star formation which
produced the bulk of the mass, and which ended  some 3~Gyr ago, a
constraint set by the age of the stars in the bar and the nucleus. Our
best-fitting model is one in which the current star formation is but
the latest of a series of relatively short bursts of star formation
which have occurred for the last 500~Myr or so.  A clear bi-polar
azimuthal age gradient is seen within
the ring, with the youngest stars occurring near where the bar dust
lanes connect with the ring.\\
\emph{Conclusions} Our kinematic and morphological results all confirm
the picture in which the nuclear ring in M100, considered typical, is
fed by gas flowing in from the disc under the action of the bar, is
slowed down near a pair of resonances, and forms significant amounts
of massive stars. Detailed stellar population modelling shows how the
underlying bulge and disc were put in place a number of Gyr ago, and
that the nuclear ring has been forming stars since about 500~Myr ago
in a stable succession of bursts. This confirms that nuclear rings of
this kind can
form under the influence of a resonant structure set up by a bar, and
proves that they are stable features of a galaxy rather than one-off
starburst events. 
\end{abstract}



\section{Introduction}
The presence of a bar or other asymmetric perturbation of the
gravitational potential within a spiral galaxy can strongly influence
the dynamics of the gas and the stars. Bars are commonly formed within
dynamically unstable galactic discs. During their evolution they can
experience vertical buckling instabilities, which thicken the bar and
increase the vertical velocity dispersion \citep{com90,raha91}, and
which we now know can be recurrent \citep{mart06}. Bars are  present
in up to 70$\%$ of spiral galaxies (Knapen, Shlosman \& Heller 2000a; Eskridge et
al. 2000; Grosb$\o$l, Patsis \& Pompei 2004), indicating that they
must be long lived components of a galaxy. This strong conclusion is
confirmed by recent results suggesting that the bar fraction is
roughly constant out to $z\sim1$ (e.g., Jogee et al. 2004). 

Dust lanes are often seen within large scale bars, offset from the bar
major axis. The main families of periodic orbits making up the bar
are elongated parallel to the bar, and perpendicular orbits are
present if one or more inner resonances exist (Contopoulos \&
Papayannopoulos 1980), so called $x_1$ and $x_2$ orbits,
respectively. At the location of such resonances, between the bar pattern speed and the period of the radial oscillations of the stars and gas, the stellar orbits abruptly change orientation. The gas orbits can only do this gradually, and this leads to the existence of shocks within the bar. The compression of this gas is manifested by increased dust extinction. 

The shocks can lead to angular momentum loss in the gas, as a result of which the gas will flow in towards the nucleus until it stalls in a ring-like region in the vicinity of the Inner Lindblad Resonances (ILRs), where there is no net torque acting on the gas (Knapen et al. 1995a, hereafter K95a; Heller and Shlosman 1996). The accumulation of gas in such a nuclear ring can easily lead to enhanced massive star formation (Heller and Shlosman 1994; K95a). Such star forming rings are found within the central 1-2 kiloparsec of 20$\%$ of spiral galaxies, practically all barred \citep{k05}, and are traced by strong line emission. Overall, nuclear rings contribute around 3-5\% to the local star formation rate \citep{ken05}, and may play a role in the secular evolution of galaxies (see review by Kormendy \& Kennicutt 2004).

The galaxy M100 (=NGC~4321) was chosen as the focus of this study due
to its relative proximity \citep[16.1~Mpc,][]{Fer96}, giving a scale of
70 parsec per arcsec, and the moderate strength of its bar. The bar
strength ($Q_{\mbox{b}}$) of M100 has been estimated at 0.2 by
Laurikainen \& Salo (2002), who suggest that significant asymmetric forces
are present for bar strengths of $Q_{\mbox{b}}> 0.05$. M100 exhibits a
particularly clear resonant circumnuclear structure which has been
described in detail by K95a and Knapen et al. (1995b, hereafter K95b,
and 2000b, hereafter K00). The morphology changes dramatically between
wavebands. In the \emph{K} band, a small-scale bar is clearly
visible, at the same position angle (PA) as the large-scale
bar (see also Section 6.3). The picture, therefore, is of a single bar dissected by a nuclear
ring. Hotspots of \emph{K} band emission are located at the ends of
the inner part of the bar. In blue optical light, as well as in
H$\alpha$ emission, a bright star-forming two-armed spiral structure
is evident, delineated by the offset dust lanes which can be traced
inwards through the main bar, through the ring, and towards the
centre. We define the contact points as the locations on the ring
dissected by a line through the nucleus and perpendicular to the bar major axis. 

In this paper, we present stellar and gas kinematics, absorption line
indices (H$\beta$, Mg$\emph{b}$ and Fe5015), and {\it Spitzer Space
  Telescope} ({\it SST}; Werner et al. 2004) near- and mid-infrared
imaging across the nuclear ring and bar region in M100. We aim to study the detailed interplay between the dynamics on the one hand, and the ancient and recent history of the star formation in the bar and ring on the other.

This is the second paper presenting results from our new dataset on this galaxy. In the first paper (Allard, Peletier \& Knapen 2005; hereafter Paper~I) we presented emission line and gas velocity dispersion maps which showed how massive star formation occurs at the precise locations of relatively cool gas, and related these results to the dynamical origin of the ring in terms of a resonance structure set up by the bar. The current paper is structured as follows: Section 2 describes the observations, data reduction and analysis of the SAURON and {\it SST} data, Section 3 briefly describes the morphology and line ratios, Section 4 concentrates on the kinematics, and Section 5 on the stellar populations. Our discussions and conclusions are presented in Sections 6 and 7.

\section{Observations, Data Reduction and Analysis}
\subsection{SAURON observations}
The central region of M100 was observed with the SAURON \citep{bac01}
integral field unit (IFU), mounted to the William Herschel Telescope
(WHT), on La Palma on 2003 May 2. To cover the whole bar region, three
separate pointings were made, the positions of which are shown in
fig.~1 of Paper~I. Each separate field of view covers an area of 33$\times$41 arcsec on the sky, which is mapped by a lenslet array, containing 1431 square lenses of 0.94$\times$0.94 arcsec in size. A further 146 lenses are available 1.9 arcmin away from the field for simultaneous observation of the sky background. After passing through the lenslet array, the light from the source enters a transmisson filter, which is tilted to prevent overlap of the spectra on the CCD array. The filter allows the wavelength range 4760-5400\,{\AA} to be observed. The large number of spatial elements in SAURON limits the wavelength coverage due to lack of space on the CCD array (EEV 12 with 2148$\times$4200 pixels). This range was chosen as it contains the stellar absorption features H{$\beta$}, Mg$\emph{b}$, some Fe{\sc i} features, and the emission lines H{$\beta$}, [O{\sc iii}] and [N{\sc i}]. The light through each lens is then dispersed by a spectrograph, giving 1577 spectra on the CCD array. Each field was exposed for 3$\times$1800\,s, and dithered by one or two lenses to avoid systematic errors. Each single exposure was taken between two comparison neon lamp exposures for accurate wavelength calibration. Standard stars were also observed for flux calibration, and continuum and sky flat exposures were taken for flat-fielding. The seeing was around one arcsec during the night and the conditions were clear but not entirely photometric. 

\subsection{SAURON data reduction}
The data were reduced, extracted and calibrated using the specially developed XSauron software \citep{bac01}. Firstly, the raw images are bias and dark subtracted using standard methods. Next, each individual spectrum must be extracted from the 2D image and placed in its correct position within a datacube. This requires the creation of a mask model, which finds the position of each individual spectrum. Datacubes of all the object, lamp, and flat field exposures are created. The neon lamp datacubes are used to calibrate the wavelength dimension, and the continuum lamp and sky flat datacubes are used to remove low frequency flat field effects. The spectral resolution of each lens may be different, due to optical distortions across the field. \citet{bac01} found that the instrumental spectral resolution can range from 90\,km\,s$^{-1}$ in the centre to 110\,km\,s$^{-1}$ in the outer regions. This had to be corrected since multiple exposures of the same object are to be combined, which may fall on different regions on the detector. The instrumental resolution is measured from the sky flat datacube, and each lens is then smoothed to the lowest resolution ($\sigma$=2.1~\AA) using a Gaussian convolution. Cosmic rays are removed using an algorithm which contains both spatial and spectral morphological criteria, the sky background is subtracted, and the object datacube is flux calibrated using the standard stars observed. The wavelength range of each lens will vary slightly across the field. This is due to the filter being tilted to avoid ghost images. To achieve a common range all the spectra are truncated to the range 4800-5380~\AA. To merge the three 1800s exposures, an image is created by integrating the total flux and collapsing each datacube along its spectral dimension. As standard stars were not taken after each galaxy exposure to save time, the datacubes are normalised to correct for small errors in the flux calibration. Using the integrated images, the datacubes can be recentred on a bright star or object within the field. The separate regions of the galaxy observed are then mosaiced to produce one large datacube. As there is only a small region of overlap between the three fields, there are no bright objects with which to match the fields together. Instead, the PAs and RA and dec of each field were obtained from the file header, and used to calculate the rotation and offsets needed to mosaic the field together and align it North-East. An H$\alpha$ image of the galaxy was used to verify that this had worked correctly.  

The final datacube has a new spatial grid with a sampling of
0.8$\times$0.8 arcsec, a spectral sampling of 1.1\,\AA ~and now
contains 5775 individual spectra. The datacube was spatially binned
using the Voronoi 2D binning method of \citet{cap03} to achieve a
minimum signal to noise ratio across the field to extract accurate
kinematics and line strengths. The data were binned to a signal to
noise ratio of 60 per resolution element which resulted in 802
bins. The bins have been left unsmoothed in all subsequent images used
in this paper.  

\begin{figure}
\begin{center}
\includegraphics[width=0.45\textwidth]{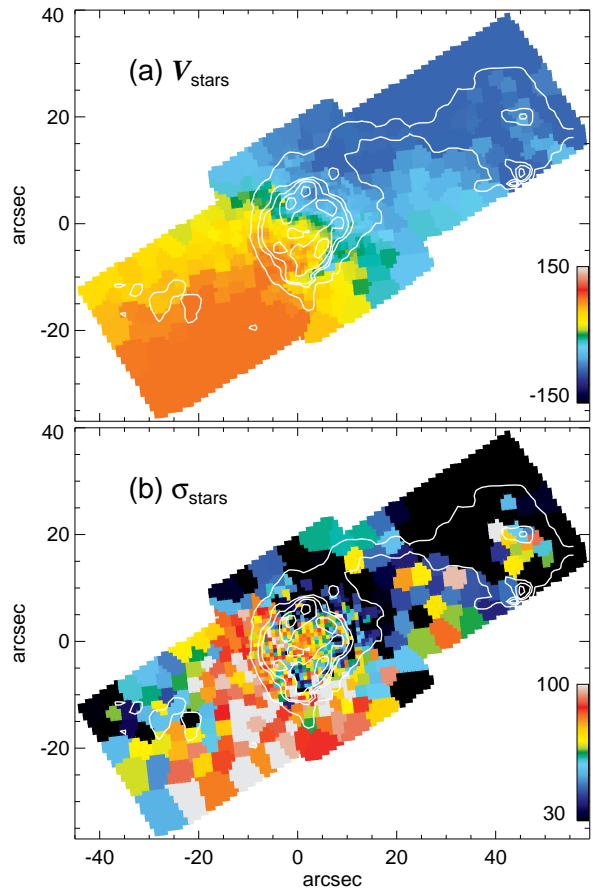}
\caption{(a). Stellar velocity field. The systemic velocity of 1571\,km\,s$^{-1}$ has been subtracted.  (b) Stellar velocity dispersion, overlaid with H$\beta$ emission contours at relative intensity levels [0.05,0.3,0.5,1,2].  
Units are in km\,s$^{-1}$. In this figure, and all following figures, North is up and East is to the left.}
\end{center}
\end{figure} 

\subsection{SAURON data analysis}

The spectrum of M100 contains contributions from each star along the line of sight. The spectral features are broadened due to the differences in velocity between the stars which were observed with each SAURON aperture. The observed spectrum is a superposition of the spectra of these stars, and, assuming they share the same kinematics, can be well described by the sum of the spectra representative of the stellar population, each convolved by the line-of-sight-velocity distribution (LOSVD). In addition to stars, gas is also observed along the line of sight, and so the spectrum of M100 additionally contains a number of strong emission lines (H$\beta$, [O{\sc iii}], [N{\sc i}]). This gas contribution needs to be removed to obtain a clean stellar spectrum before any absorption line strengths can be determined.

The SAURON stellar kinematics of M100 were obtained by the fitting of its spectra with a linear combination of single-age, single-metallicity population models from \citet{vaz99} convolved with a Gauss-Hermite parametrization of the stellar LOSVD (Gerhard 1993; van der Marel \& Franx 1993). The best-fitting parameters were determined using the Penalized Pixel Fitting (pPXF) method of \citet{cap04}, which allows the exclusion of the regions containing strong emission lines. 

The adopted template library contains 48 models with ages ranging from 250~Myr to 17.78~Gyr, and metallicities ranging from $-$1.68 to $+$0.20 [Fe/H]. This library is similar to that used by \citet{fal06} and \citet{gan06} for the analysis of spiral galaxies. 

The procedure to measure the gas kinematics has been refined since Paper I. We now follow the new procedure as described in \citet{sar06} in which the gas emission lines are treated as additional Gaussian templates, which are fitted to the data simultaneously with the stellar population models. Furthermore, the fitting of the emission lines has been performed unconstrained, allowing the derivation of separate [O{\sc iii}] and H$\beta$ kinematics. Our previously determined gas kinematics were derived by fitting the emission lines together, and as H$\beta$ is the dominant line over most of the field, the gas kinematics of Paper I are closely matched by the H$\beta$ kinematics presented in this paper.

Once the gas contribution has been successfully estimated, it can be removed from the original galaxy spectrum to produce a clean, emission free stellar spectrum, and this is repeated for each spectrum in the datacube. This cleaned datacube is used for the measurements of absorption line strengths.

\subsection{Instrumental dispersion issues}
The stellar dispersion map (Fig.~1b) shows an unusual area of high
dispersion offset toward the South-West of the centre by
$\Delta$(RA)$=-10$\,arcsec, $\Delta(\delta)=-15$\,arcsec. This area is
also visible to a lesser extent in the dispersion maps for the gas
emission lines. An effect such as this could be produced by an
inaccurate homogenisation of the spectral resolution across the field,
so this step in the data reduction process was investigated more
carefully. It should be noted that this data reduction process is
identical to that used for other SAURON papers
\citep{bac01,dezeeuw02,em04,sar06,mcderm06,fal06,kun06,cap06} and no
issues of this kind have been seen before. To check whether the
correction made was adequate, the same correction was applied to one of the lamp datacubes for each of the three positions, and the dispersion of the lamp lines was then measured. A residual gradient was found of around 0.2~\AA, or 10~km\,s$^{-1}$, in each of the three fields. A correction map was created which was applied to the stellar dispersion map, but the problem area remained. Not only was this residual dispersion not large enough to account for the observed region of high dispersion, it affected the three fields in the same way, whereas the high dispersion region is predominantly in the Eastern field. 
As all the observations were taken on the same night without any large
movement of the CCD and instrument it is unlikely that the resolution of the lens changed between observations. Furthermore we used lamp datacubes instead of the sky flat datacube to compute the spectral resolution correction for each corresponding object datacube, but the high dispersion region remained in each case. 
We also checked whether the proximity of the sky lenses to the galaxy could cause strange features within the object spectra, and found that this was not the case. \citet{bat05} presented IFU observations of the stellar kinematics of M100 for the central 33$\times$29 arcsec. Their stellar velocity field agrees very well with ours, but their maps show no sign of this high dispersion region. 

We conclude that, even though the stellar velocity dispersion on the Eastern side of the centre might appear too high, we cannot identify an observational or instrumental cause. It is important to notice, however, that whatever causes this effect will only impact on the stellar and gas dispersion maps, and not the gas fluxes, or the line ratios, as these measurements are not dependent on the width of the spectral features. The line strength measurements are affected, since they are corrected for velocity dispersion (see Section 5.1), but the corrections are small ($<$ 0.2\,\AA) since the velocity dispersions are $<$100\,km\,s$^{-1}$ (see, e.g., Davies, Sadler \& Peletier 1993).

\begin{figure*}
\begin{minipage}{180mm}\includegraphics[width=0.9\textwidth]{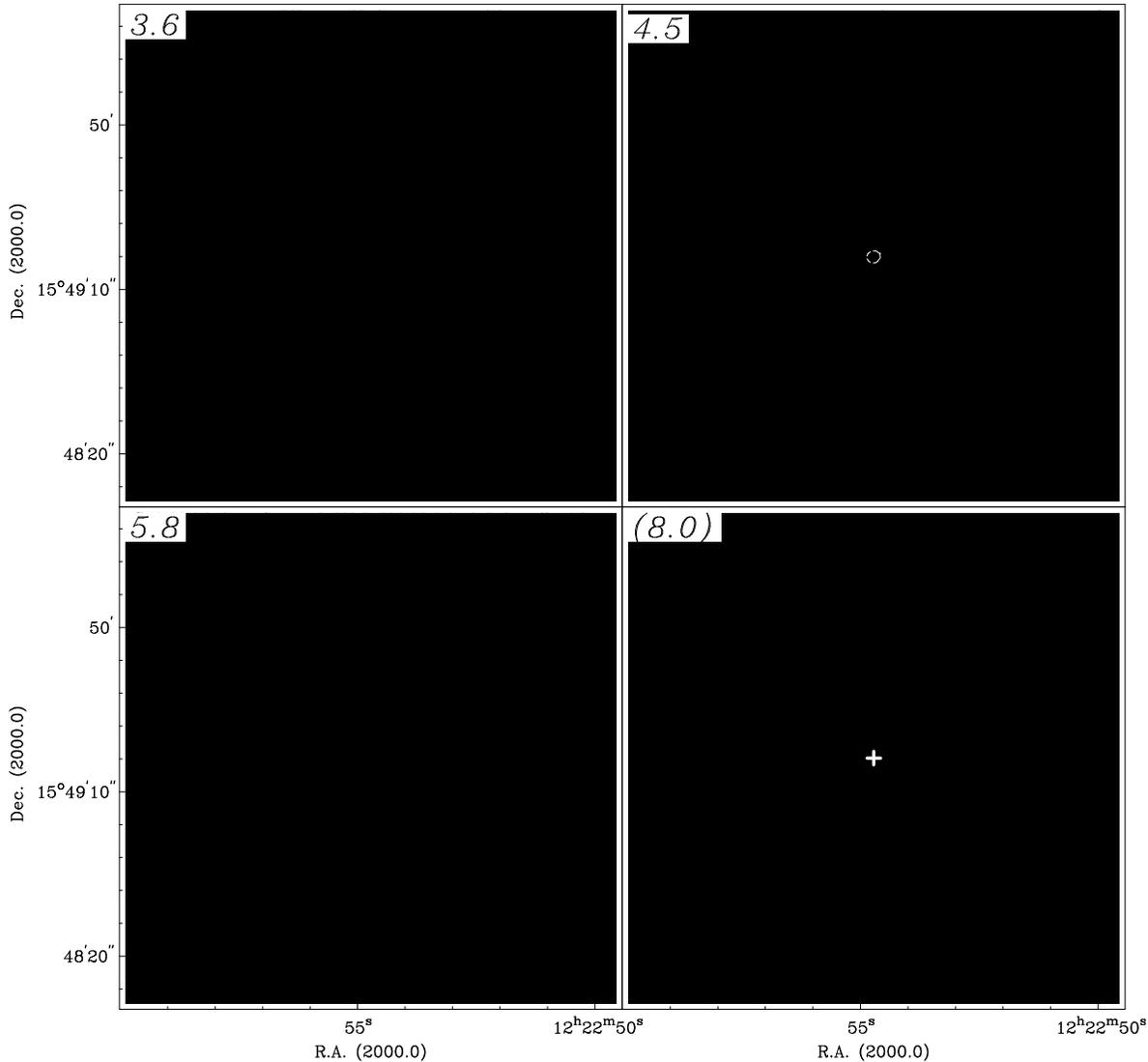}
\caption{{\it SST} IRAC 3.6, 4.5, and 5.8\,$\mu$m images of the central
region of M100, as well as ({\it bottom right}) the dust-only 8\,$\mu$m
image (Sect.~2.5). Contours are plotted logarithmically
at 0.63, 1, 1.6, 2.5, 4, 6.3, 10, 15.8, 25.1, and 39.8\,MJy\,sr$^{-1}$
for the 3.6, 4.5, and 5.8\,$\mu$m images, with an additional contour
at 63.1\,MJy\,sr$^{-1}$ for the 3.6 and 5.8\,$\mu$m images. The
nucleus of the galaxy is indicated with a cross.
}
\end{minipage}
\end{figure*}

\begin{figure}
\includegraphics[width=0.45\textwidth]{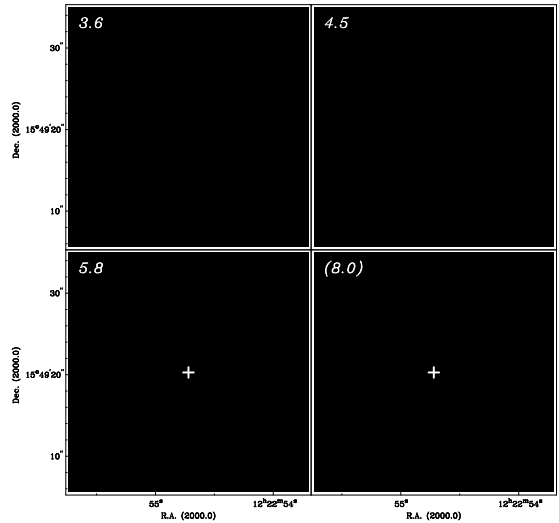}
\caption{{\it SST} IRAC images as in Fig.~2, but now highlighting the
structure of the nuclear ring region.}
\end{figure}

\begin{figure*}
\begin{minipage}{180mm}
\includegraphics[width=0.9\textwidth]{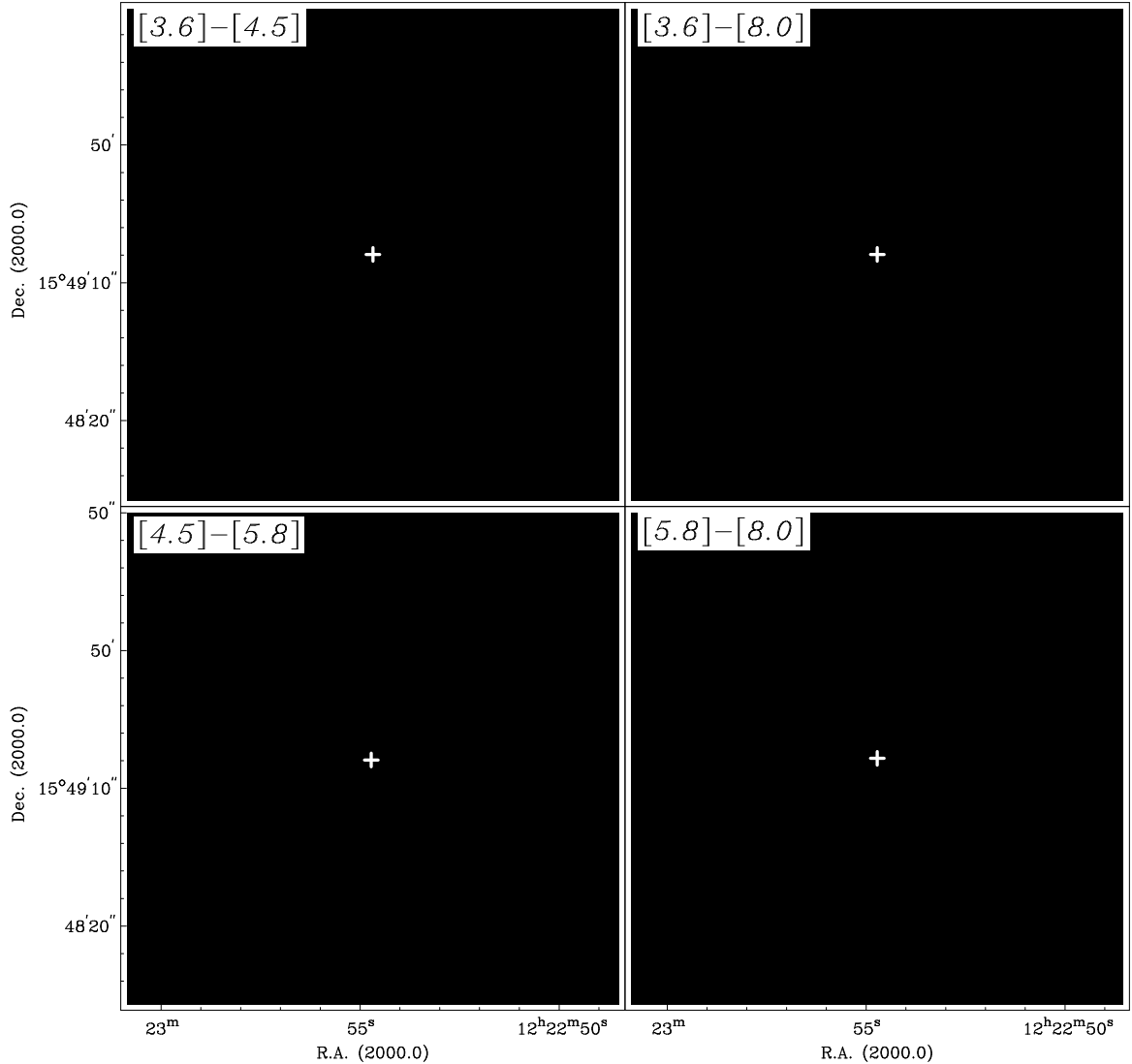}
\caption{{\it SST} IRAC colour index  maps of the central region of
  M100. In all maps, blue tones
indicate blue colours (e.g., more 3.6 than 4.5\,$\mu$m emission), while
on the other extreme of the range red and white tones indicate the
reddest colours. The nucleus of the galaxy is indicated with a white
cross.
}
\end{minipage}
\end{figure*}

\begin{figure}
\includegraphics[width=0.45\textwidth]{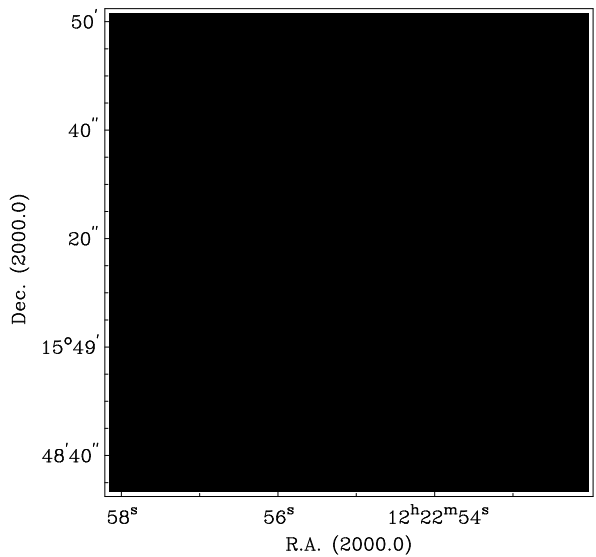}
\caption{{\it SST} MIPS 24\,$\mu$m image of the central region of M100.}
\end{figure}

\subsection{{\it SST} imaging}
We obtained {\it SST} Infrared Array Camera (IRAC; Fazio et al. 2004)
and Multiband Imaging Photometer for {\it Spitzer} (MIPS; Rieke et
al. 2004)  imaging data of M100 from the
archive, data which were obtained as part of the SINGS project
(Kennicutt et al. 2003). For each of the bands, we combined the two
sets of data taken. We refer to Kennicutt et al. (2003) and subsequent
papers by the SINGS team for all details of the observations and data
reduction. We subtracted residual sky backgrounds, rotated the images
to place them in a N-E configuration to be compared directly with our
other data, and produced colour index images for the IRAC data. The latter
were made in magnitudes, by taking the logarithm of the ratio of two
images and multiplying that with 2.5. Following the prescription of Pahre et al. (2004) and Calzetti et al. (2005) we produced a ``dust emission" image at 8\,$\mu$m by subtracting the 3.6\,$\mu$m stellar image. Pixel scales are 1.2~arcsec for IRAC and 2.45~arcsec for MIPS 24\,$\mu$m. The spatial
resolution is just over 2 pixels in each band.  We present IRAC images
in the 3.6, 4.5, 5.8 and dust-only 8.0\,$\mu$m bands of the bar region of M100
in Fig.~2, and of the nuclear ring region in Fig.~3. Colour index images of the bar region are shown in Fig.~4.
The MIPS images have too low spatial resolution to be of much use in our
study of the nuclear ring region of M100, however, we do make
reference to the 24\,$\mu$m image in Section~6.5 and hence show it in
Fig.~5.
\begin{figure}
\includegraphics[width=0.45\textwidth]{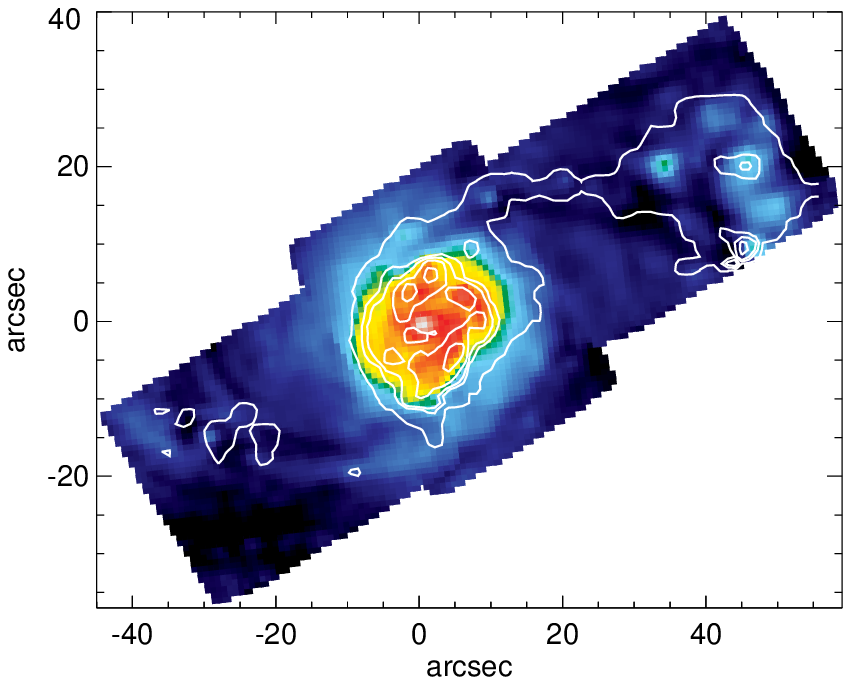}
\caption{Total intensity map, reconstructed from our SAURON datacubes. Overlaid in white are contours of the H$\beta$ emission of Figure~1.}
\end{figure}

\begin{figure}
\includegraphics[width=0.45\textwidth]{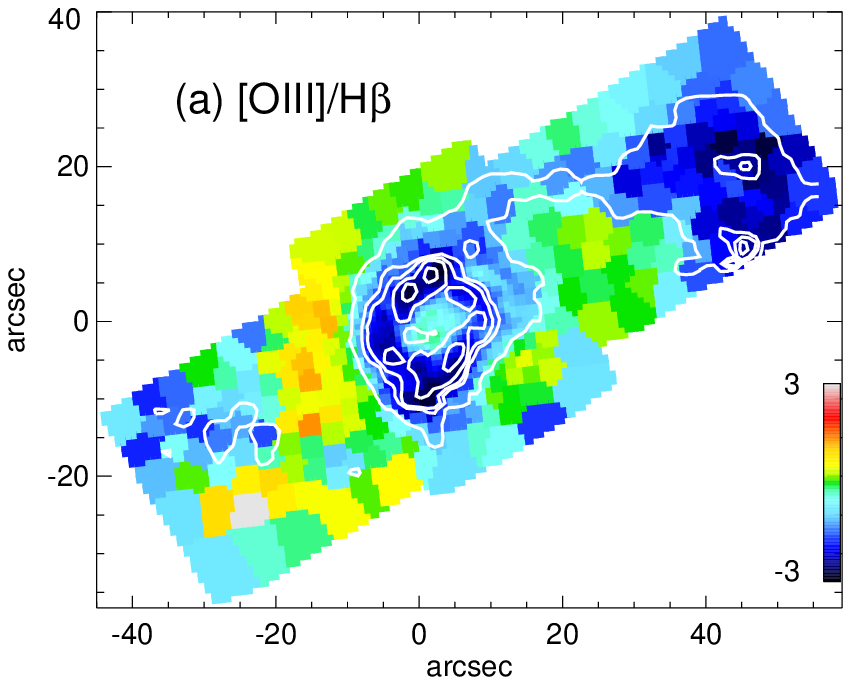}
\caption{[OIII]/H$\beta$ line ratio map, overlaid with the H$\beta$ emission contours of Figure~1. The scale is logarithmic.}
\end{figure}

\section{Morphology and Line Ratios}
\subsection{SAURON morphology}
The SAURON fields were chosen to cover the complete bar of M100, found
at a PA  of 110$^{\circ}$ (K95b). The positions of the three fields are shown in fig.~1f of Paper I. The reconstructed total flux within the observed spectral range is shown in Figure 6. The bright spiral armlets are visible, along with the offset dust lanes. 
The observed morphology in the two brightest emission lines, H$\beta$ and [O{\sc iii}], can be seen in fig.~1a and 1b of Paper I and is summarised here. Contours of the H$\beta$ emission are shown in various figures in the present paper, e.g., Figure~6. 
The H$\beta$ emission follows closely the H$\alpha$ morphology (K95b, K00). The emission is brightest in a ring of emission surrounding the centre, and at the Western end of the large bar. A thin arc of emitting material connects the ring with both ends of the bar, more strongly on the Western side. 
The [O{\sc iii}] emission is strongly centrally peaked, most likely caused by the LINER nucleus. The emission also faintly traces the H$\beta$ emission.

\subsection{Line ratios}
An emission line ratio map of [O{\sc iii}]/H$\beta$ is presented in Figure~7. A line ratio containing one heavy element and one hydrogen Balmer line is relatively insensitive to changes in abundance \citep{vei87}, and helps indicate the ionisation mechanism powering the lines. Photoionisation of gas caused by non-thermal or power-law type continuum radiation results in [O{\sc iii}]/H$\beta >$~3. Lower values of [O{\sc iii}]/H$\beta$ are suggestive of star formation, and photoionisation by OB type stars. The metallicity of the gas can also affect the [O{\sc iii}]/H$\beta$ ratio, but as metallicity does not change abruptly within a single galaxy, large fluctuations in [O{\sc iii}]/H$\beta$ within a galaxy are likely to be caused by changes in the ionisation mechanism \citep{sar06}. 

Figure~7 shows the [O{\sc iii}]/H$\beta$ ratio across the inner region of M100. The nuclear ring and the emission-line regions on the Eastern side of the
bar contain much lower values (less than [O{\sc iii}]/H$\beta=0.1$) than the
rest of the galaxy, confirming the presence of massive star formation. Within the nucleus, [O{\sc iii}]/H$\beta \sim 1$, a value typical of LINER galaxies \citep{vei87}.

\subsection{Near- and Mid-Infrared Morphology}
The overall morphology of the {\it SST} IRAC images describes emission predominantly from the stellar photospheres of the old population, unhindered by dust
extinction, and,
at the longer wavelengths, by tiny dust grains and by the large
molecules known collectively as polycyclic aromatic hydrocarbons
(PAHs). The colour index images highlight the relative differences
between the different bands.
The distribution of the old stellar population is seen at 3.6 and 4.5
\,$\mu$m. The main bar of M100, of relatively modest strength, is seen
clearly, the spiral arms in the disc are outlined, and the
star-forming regions at the ends of the main bar are relatively
modest. The most obvious feature in all bands in Fig.~2 is the strong
emission from the circumnuclear region, in which the emission from the
inner part of the bar (K95a) dominates (shown in detail in Figure~3).

The IRAC colour index images (Fig.~4) show how the
emission varies with wavelength. The nucleus is consistently `blue',
and the nuclear ring, as well as the disc spiral arms, predominantly
`red'. The [3.6]$-$[4.5] image highlights the most intense
star-forming regions, presumably because of the emission by hot dust
which is more prominent at 4.5 than at 3.6\,$\mu$m. The colour index
images involving the 5.8 and 8\,$\mu$m emission show how they trace the star-forming regions through the emission from small dust grains and PAHs. We will discuss the implications of the {\it SST} imaging on our understanding of the central region of M100 in Section 6.

\begin{figure}
\includegraphics[width=0.45\textwidth]{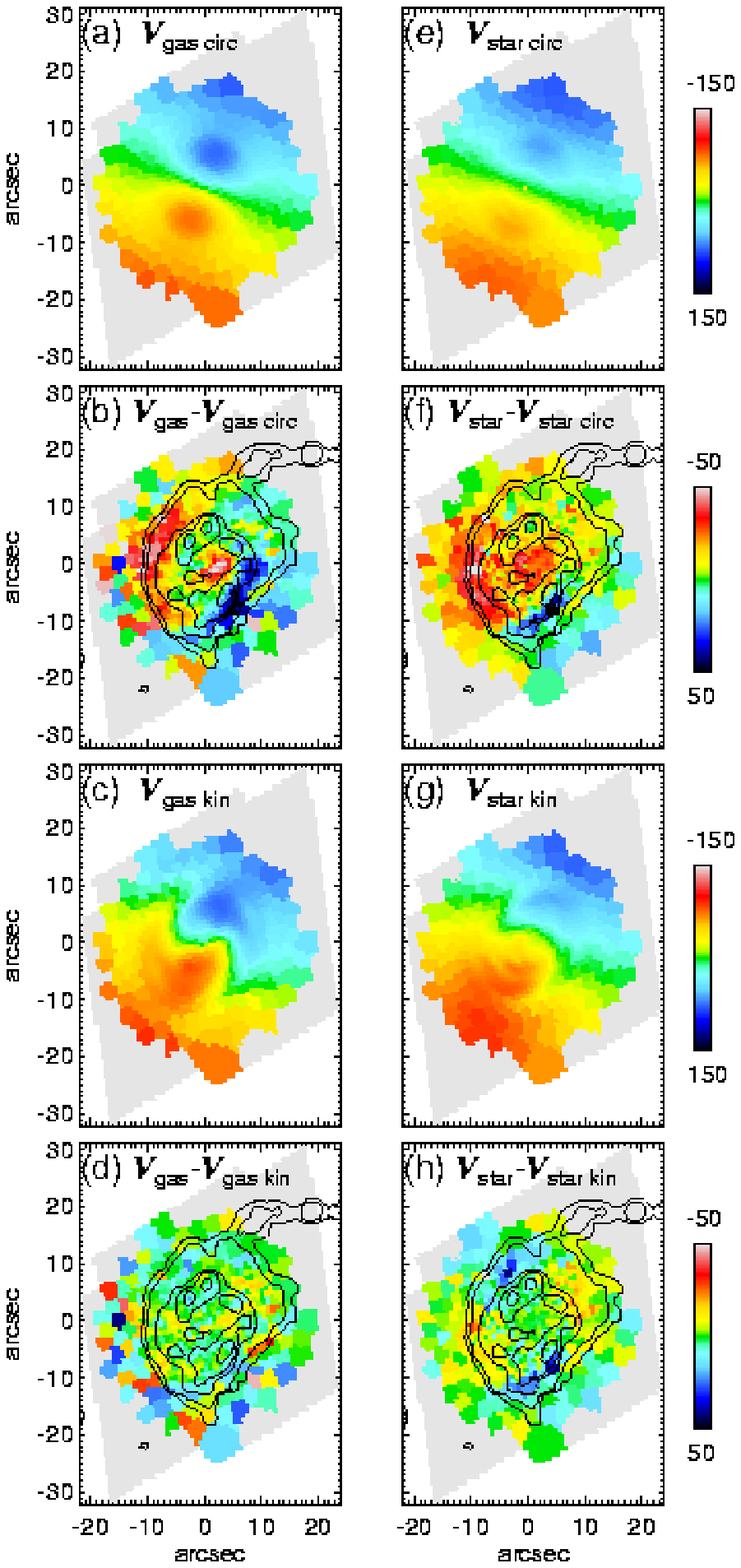}
\caption{(a). Model circular velocity field of the gas, for the region
  sampled, derived using only the first coefficient of the harmonic
  expansion. This is a simple model of an inclined rotating
  disc. (b). The residuals from the H$\beta$ gas velocity field minus
  the model circular velocity field.  (c). Model reconstructed
  velocity field of the gas, using the first five coefficients of the
  harmonic expansion. (d). The residuals from the H$\beta$ gas
  velocity field minus the reconstructed gas velocity field. (e, f, g,
  h) As above, but now for the stars.}
\end{figure}

 \begin{figure*}
\begin{minipage}{170mm}
\includegraphics[width=1\textwidth]{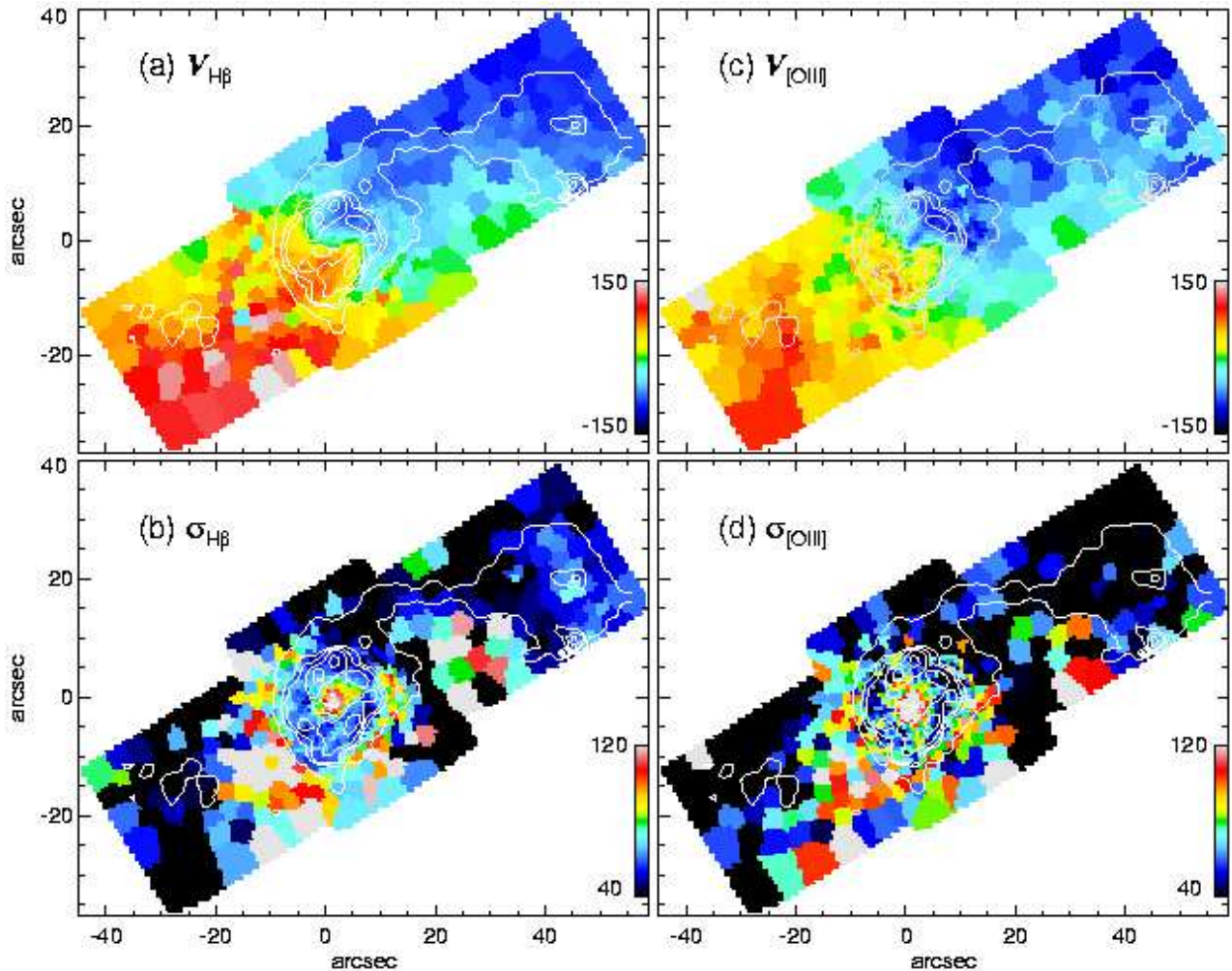}
\caption{(a). H$\beta$ gas velocity field. (b). H$\beta$ gas velocity
  dispersion field. (c). [O{\sc iii}] gas velocity field. (d). [O{\sc
      iii}] gas velocity dispersion field. Units are in km\,s$^{-1}$. The systemic velocity of 1571\,km\,s$^{-1}$ has been subtracted from velocity maps. Overlaid are H$\beta$ emission line intensity contours, as in Figure~1.}
\end{minipage}
\end{figure*}

\begin{figure}
\includegraphics[width=0.45\textwidth]{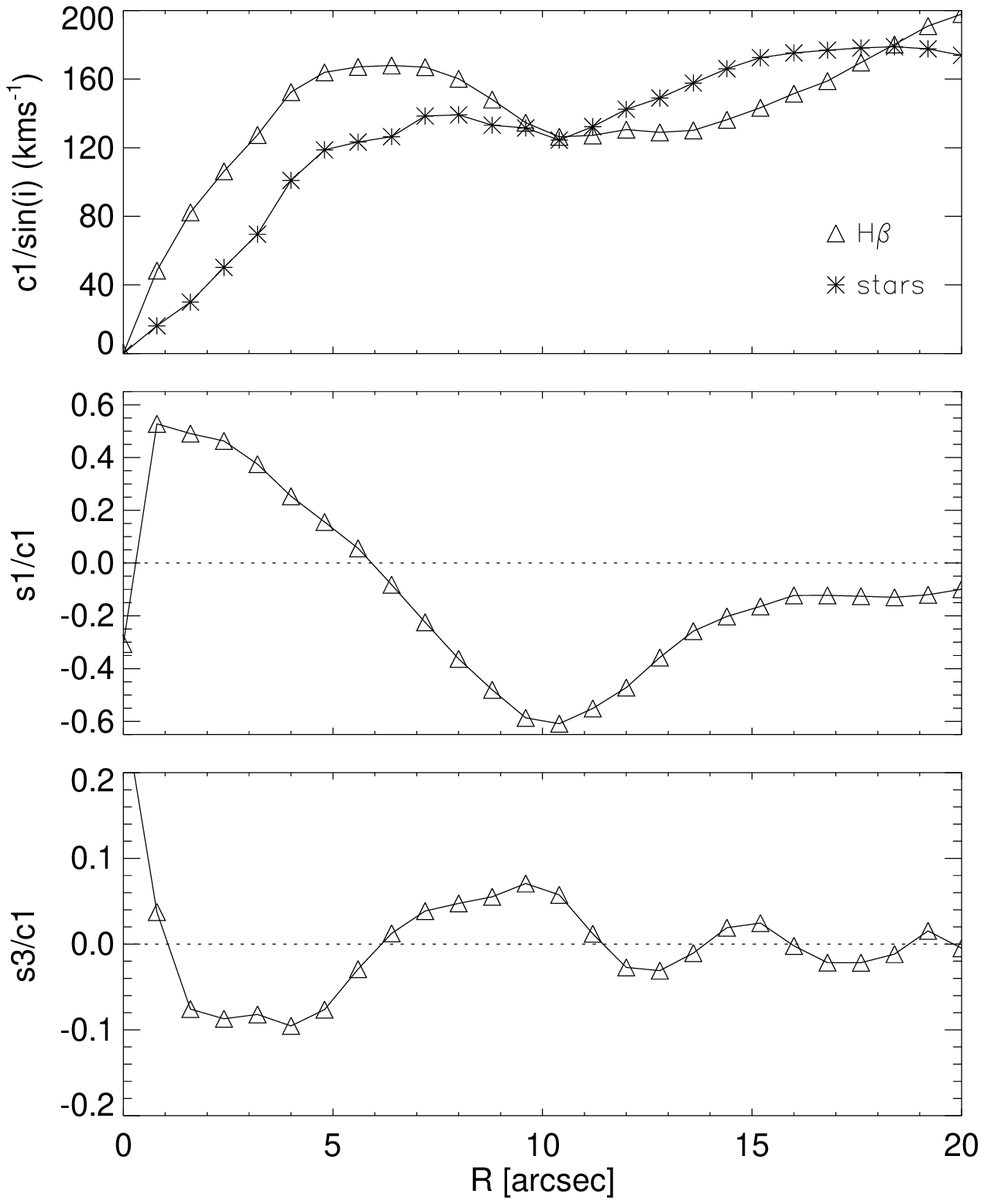}
\caption{Coefficients of the harmonic expansion against radius for the gas. Upper Panel: Circular velocity ($c_{1}/\sin{i}$, for both the gas and the stars, in km\,s$^{-1}$). Middle Panel: $s_{1}/c_{1}$. Lower panel: $s_{3}/c_{1}$.}
\end{figure}

\section{Kinematics}
The kinematics of the nuclear ring region in M100 has been studied in detail by K00, who used observations in the H$\alpha$ and CO $J=1\rightarrow 0$ lines, and compared them to the numerical model of K95a. Our SAURON data allow us to extend the kinematic analysis by studying the stellar as well as the gas kinematics over a larger spatial area, and by using a more quantitative technique to investigate the non-circular motions in the ring region. 

We have used the technique known as kinemetry (Copin 2002; Krajnovi\'c
et al. 2006) to analyse the stellar and gas velocity fields quantitatively. Kinemetry involves a procedure similar to surface photometry, but for the non-zero order moments of the LOSVD. Best-fitting ellipses are determined across the surface of the velocity maps, along which the profiles of the moments can be extracted and analysed by means of harmonic expansion. 
In photometry, galaxy light can be described by an ellipse, with any deviations from this giving rise to extra terms in the Fourier expansion. These terms represent quantities such as the diskiness or boxiness of the ellipse. When using kinemetry to analyse velocity maps, these extra terms carry information on how the velocity field deviates from circular motion viewed for a given inclination. Each fitted ellipse is defined by two parameters, PA and flattening, \emph{q}, where $q$ is related to the ellipticity $e$ by $q=1-e$, and to the inclination $i$ (assuming a thin disc) by $q= \cos(i)$. We have fixed the PA at 153$^\circ$ and the inclination at 27$^\circ$ (Knapen et al. 1993, K00). 

The velocity profiles along sampling ellipses in galaxies are well represented by a cosine law (Krajnovi\'c et al. 2006). A velocity map $V(r,\theta)$ can be divided into elliptical rings with velocity profiles described by a number of harmonic terms: 
\[
V(r,\theta)=c_{0}(r)+\sum_{n=1}^{N} s_{n}(r) \sin(n\theta)+c_{n}(r)\cos(n\theta)
\]
where $(r,\theta)$ are the polar coordinates of the velocity field. In the case of perfect circular motion, $c_{0}=V_{\mbox{sys}}$ and $c_{1}=V_{\mbox{c}}\sin i$, with all other coefficients equal to zero. The line of sight velocities can then be represented by the following:
\[
V(r,\theta)=V_{\mbox{sys}}+ V_{\mbox{c}}(r)\cos(\theta)\sin(i) 
\]  The subtraction of a disc model of circular motion from the observed velocity field then reveals the non-circular motions.

In the case of velocity fields containing non-circular motions, the higher order coefficients $c_{n}$ and  $s_{n}$ become non-zero. Franx, van Gorkom \& de Zeeuw (1994), Schoenmakers, Franx \& de Zeeuw (1997) and Wong, Blitz \& Bosma (2004) have investigated the behaviour of these coefficients for a number of disc models. It is possible to discriminate between various models of non-circular motions (e.g., radial flows, warps, elliptical streaming) by an inspection of the $c_{1}$, $s_{1}$, and $s_{3}$ terms. The coefficient $c_{1}$ is always dominant, and relates to the amplitude of the rotation curve. All other coefficients are commonly expressed as fractions of $c_{1}$. The simplest interpretation of the $s_{1}$ term is as a radial flow, with its sign indicating inflow ($-$) or outflow ($+$). The $s_{3}$ term reveals asymmetric structures in the disc, where a constant non-zero $s_{3}$ is a signature of a global, elongated potential such as a bar, whereas fluctuations in $s_{3}$ reveal localised structures such as spiral arms. 

The velocity field reconstructed from the derived coefficients will reproduce the overall rotation and non-circular motions of the observed velocity fields. We present such reconstructed maps in Fig.~8, where we used all components in Fig.~8c and 8g and only component $c_1$ in Fig.~8a and 8e (circular rotation only). Subtracting these reconstructed maps from the observed velocity field yields the residual maps shown in Fig.~8d, 8h, 8b and 8f, respectively. The analysis is restricted to the central 60$\times$40 arcsec because only for this area the velocity field is fully sampled at a given radius.

\subsection{Gas kinematics}

The gas velocity fields of [O{\sc iii}] and H$\beta$ (Fig.~9) both show slightly faster rotation than the stars, and are more complicated. H$\beta$ rotates slightly faster than [O{\sc iii}] at larger radii. In the central 30 arcsec diameter region which contains the nuclear ring, K00 reported two distinct sets of features in the H$\alpha$ velocity field which they related to non-circular motions due to gas streaming around the nuclear part of the bar and the spiral arms just outside the nuclear ring. We confirm the existence of these features in our gas velocity fields. Firstly, we see the distinct S-shaped deviation due to the inner part of the bar in the velocity contours of H$\beta$, at a radius of $\sim 5$ arcsec, which is also apparent but less clear in the [O{\sc iii}] velocities. Secondly, at a radius of $\sim 9$ arcsec we see the spiral density wave streaming motion as a set of deviations from circular motion, but in the opposite sense as those due to the inner part of the bar.
Similar effects on the iso-velocity contours are regularly seen in the central regions of other galaxies observed using IFUs (e.g., Hernandez et al. 2005; Ganda et al. 2006; Falc\'on-Barroso et al. 2006; Sarzi et al. 2006). 

K00 used their kinematic observations to confirm the predictions from the three-dimensional numerical modelling of K95a. This model required the existence of two ILRs in the region, with the nuclear ring lying at the inner ILR. They explained the origin of the non-circular motions firstly, as gas streaming along the inner part of the bar, and secondly, as the effects of a bar-induced spiral density wave. 

In Paper~I we presented a map of the combined H$\beta$ and [O{\sc iii}] gas velocity dispersion, which shows a ring of lower dispersion gas, lying at the exact position of the nuclear ring as seen in H$\beta$ emission. We suggested that massive stars are forming from the relatively cool gas which flows in from the disc through the bar (Paper~I), consistent with the very low values of [O{\sc iii}]/H$\beta$ seen in the ring. In Fig.~9, we show the gas velocity dispersion separately for H$\beta$ and [O{\sc iii}]. These maps obviously show the same structure as the combined map presented in Paper I, albeit with a lower signal to noise ratio. Also, the [O{\sc iii}] gas velocity dispersion does not reach values as low as H$\beta$ within the ring, and has a much higher value in the centre. This can be understood considering that the [O{\sc iii}] lines are not as reliable as H$\beta$ to trace the denser and dynamically colder components of the interstellar medium (ISM), where star formation is more likely to occur. Indeed sources other than ionisation by O-stars can contribute efficiently to the production of [O{\sc iii}], such as shocks and AGB stars. 
 
Figure~10 shows the results from the kinemetric analysis for the
gas. Both H$\beta$ and [O{\sc iii}] gave similar results, so we
present the full kinemetric analysis for H$\beta$ only. The rotation curve
for the gas rises quickly, to a value of 135~km\,s$^{-1}$ within a
radius of 5\,arcsec (Figure~10). The velocities then flatten out to
around 150~km\,s$^{-1}$, up to 20\,arcsec. The combination of the large
offset between the PA  of the bar (110$^\circ$) and that of the kinematic axis (153$^\circ$) with the footprint of our three combined SAURON pointings means that ellipses can only be fitted out to a radius of 20 arcsec. The general shape of the gas rotation curve is similar to that derived by K00 using H$\alpha$ Fabry-P\'erot data. The initial gradient within the H$\alpha$ data is much steeper due to the increased spatial and spectral resolution, but at larger radii the two curves are comparable.

The higher order coefficients $s_{1}$ and  $s_{3}$ are non-zero
(Fig.~10), revealing the non-circular contributions to the velocity
profiles. Wong et al. (2004) performed a kinematic analysis of M100
using CO and H{\sc i} data for the outer 20 to 200 arcsec, and our
data from 0 to 20 arcsec match up well with theirs. The $s_{1}/c_{1}$
graph is positive within the inner 5~arcsec, and then becomes negative
for the rest of the radial range. The most negative values lie at the
location of the nuclear ring at $\sim 10$ arcsec. This may indicate
radial inflow at this location. The $s_{3}/c_{1}$ graph shows a number
of small fluctuations within the 20~arcsec range, with a positive peak
at $\sim 9$ arcsec and smaller peak at at $\sim 15$ arcsec, which
correspond to the locations of the nuclear ring and spiral arms,
respectively, indicating that these structures are responsible for localised perturbations in the gravitational potential. 

The residual velocity field (Fig.~8b) shows large regions of non-circular motions in the gas. Along the bar minor axis, the sign of the residual changes twice, indicating the two sets of streaming features mentioned above. The residuals of the observed velocity field minus the reconstructed velocity field (Fig.~8d) are a measure of the accuracy of the kinematic fit. The residuals are small except for two regions along the bar major axis, where the model does not quite match the steep initial gradient in the velocity field. 

\subsection{Stellar kinematics}
The stellar kinematics of M100 across the bar and central region are presented in Figure~1. The stellar velocity field shows mostly regular rotation, with velocities increasing with radius until, just outside the nuclear ring, they reach a constant velocity of $\pm\sim$150 km\,s$^{-1}$. The field is mostly smooth and typical of disc rotation. We can recognise the two sets of features described above for the gas kinematics, which are due to bar and density wave streaming. This proves, for the first time, that the dynamics giving rise to the observed non-circular motions in the gas affects the stars to a very similar extent, contrary to many other spirals (see, e.g., Falc\'on-Barroso et al. 2006; Ganda et al. 2006). 

In comparison to the gas kinematics, the stellar kinematics in M100 have received very little attention in the literature. Batcheldor et al. (2005) present the stellar velocity field of M100 for the central 33$\times$29 arcsec, which agrees well with ours, with similar rotational velocities and non-circular motions. 
Barth, Ho \& Sargent (2002) measured a central velocity dispersion of
92$\pm$4 km\,s$^{-1}$ from their red spectra using an aperture of 3.74
arcsec, while a value\footnote{This value differs from that given in
  Paper I, where we give 98$\pm$4 km\,s$^{-1}$ due to a correction
  applied to the stellar dispersion map (see Section 2.4), but this would not affect any of the conclusions of Paper I.} derived from a similar aperture on our data yields 85$\pm$4 km\,s$^{-1}$.
                                                         
The kinemetry analysis described in  Section 4.1 was repeated for the stars. The higher orders of the harmonic coefficients are not shown as they are difficult to interpret in the case of the stars. The stellar rotation curve, derived here for the first time, is different in details from that of the gas. The stellar curve has a shallower initial gradient, rising to around 110~km\,s$^{-1}$ within 7 arcsec. This value then stays constant until around 12 arcsec, when the velocities rise again to become comparable with the gas velocities.

The residuals between the observed velocity field and the circular
velocity and reconstructed velocity fields are shown in the right side
of Figure~8. Figure~8f shows a similar structure to that of the gas
(Fig.~8b) but less extreme. This similarity could be the kinematical signature of young stars forming within the inflowing gas. The residuals shown in Fig.~8h are small (less than 10~km\,s$^{-1}$), indicating that our assumption when applying kinemetry that the motions can be fitted to zero-th order by an inclined rotating disc is justified.

\section{Stellar Populations}
\subsection{Deriving line strengths}
The combination of the wavelength range of SAURON and the systemic velocity of M100 allow the measurement of the H$\beta$, Mg{\emph b} and Fe5015 indices, as defined on the Lick/IDS system (Worthey 1994). The line indices are measured from the spectrum after the gas emission contribution has been removed. The use of indices to measure absorption line strengths on this system requires the definition of a band containing the feature and two side bands containing the red and blue continuum levels. The mean level in each of the two side bands is determined and a linear relation is fitted through the midpoints to define the continuum. The difference between this continuum and the observed spectrum within the feature band provides the index value.

In order to compare indices with other published data and model predictions, the index measurements need to be calibrated to the Lick/IDS system, which involves three steps. Firstly, the observed spectra need to be degraded in resolution to match that of the Lick/IDS system, from a FWHM of 2.1\,\AA\ to 8.4\,\AA. Secondly, the LOSVD will broaden spectral features, and so will affect observed line strength measurements compared to intrinsic values. A correction factor is determined for each spectrum individually from the optimal template. Finally, the original Lick/IDS spectra were not flux calibrated, and so small offsets are introduced by differences in continuum shape. These offsets, as well as details of the methods stated here, can be found in Kuntschner et al. (2006).  

The final selection of the 48 templates in the library used to derive
the stellar and gaseous kinematics, and which is described in Section 2.3,
was decided on by judging the quality of the fit to the continuum in a
number of lenses across the field.

\subsection{Absorption line maps}

Maps of the H$\beta$, Mg{\emph b} and Fe5015 stellar absorption
indices are presented in Figure~11. These indices change over the
lifetime of a star, and can thus be used as age indicators for
integrated stellar populations. They are, however,  affected to a
similar degree by variations in metallicity. In subsequent sections,
we will interpret the absorption line measurements in detail by
comparing them with stellar population modelling, but we will briefly
describe the general morphology of the absorption line maps before
doing so.

The H$\beta$ absorption map shows an incomplete ring. The missing parts of the ring are where the strongest H$\beta$ emission is found, near the contact points between the dust lanes and the ring. The two maxima in H$\beta$ absorption index are $\sim4$ in value and lie on opposite sides of the ring along the bar major axis.  
The Mg$\emph{b}$ absorption map shows a much lower value within the
nuclear star forming ring, which we will quantify and model in the
following section. The Mg$\emph{b}$ index is anti-correlated with
H$\beta$ emission across the whole map. The lowest values are found at the contact point between the dust lane and the ring, where the H$\beta$ emission is strongest. 
The Fe5015 absorption map shows similar structure to that of
Mg$\emph{b}$. A ring can be seen with the lowest values occurring at the same position as strong H$\beta$ emission. 

\begin{figure}
\includegraphics[width=0.45\textwidth]{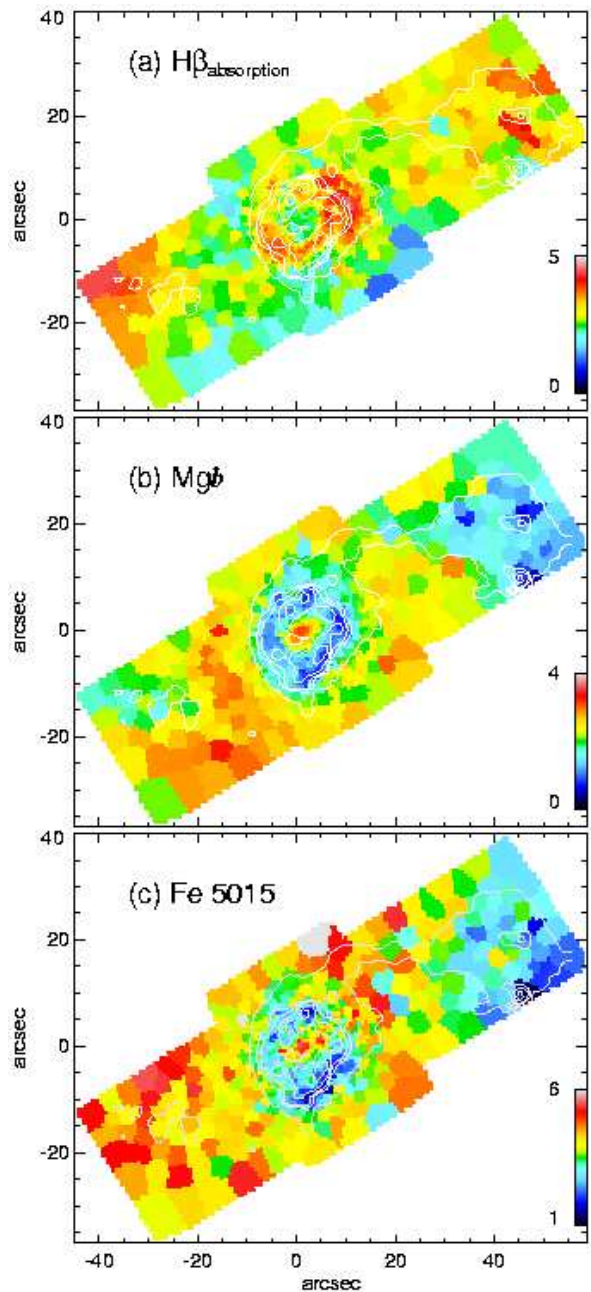}
\caption{Absorption line strength maps: (a). H$\beta$, (b). Mg{\emph b}, (c). Fe5015. Contours as in Figure~1. Units are Angstroms.}
\end{figure}

\subsection{Comparisons to model predictions: H$\beta$ versus MgFe50}
Using the three absorption line indices we can place age and metallicity constraints on the stellar populations by using the model predictions of Bruzual \& Charlot (2003, hereafter BC03). Typically, metallicity and age affect galaxy spectra in similar ways, but we can combine our available indices to break this degeneracy. The H$\beta$ index is most sensitive to the age of a population \citep{wor94}, with the strongest H$\beta$ absorption occurring in stars of around 250~Myr (see Figure~12). The Mg{\emph b} and Fe5015 indices are more sensitive to metallicity effects. \cite{fal02} defined a new index, MgFe50$= \sqrt{\mbox{Mg{\emph b}}\cdot\mbox{Fe5015}}$ which minimises the effect of the differing abundance ratios across the galaxy (see Section 5.4), and which we will use in conjunction with H$\beta$.
Circular apertures of 2 arcsec in radius were taken at points on the
ring (squares in Fig.~12) and along the bar (triangles), as well as at
the nucleus (diamond). These locations are shown, relative to the
morphology of the H$\beta$ emission, in the inset of Figure~12 (the
location of hotspots K1/2 and H$\alpha$3/4 as defined in K95a are
shown in Figure~22). The errors on the absorption line strengths,
indicated in the bottom left corner of most figures for this section, 
are Poisson errors within our apertures.

\begin{figure*}
\begin{minipage}{180mm}
\includegraphics[width=0.95\textwidth]{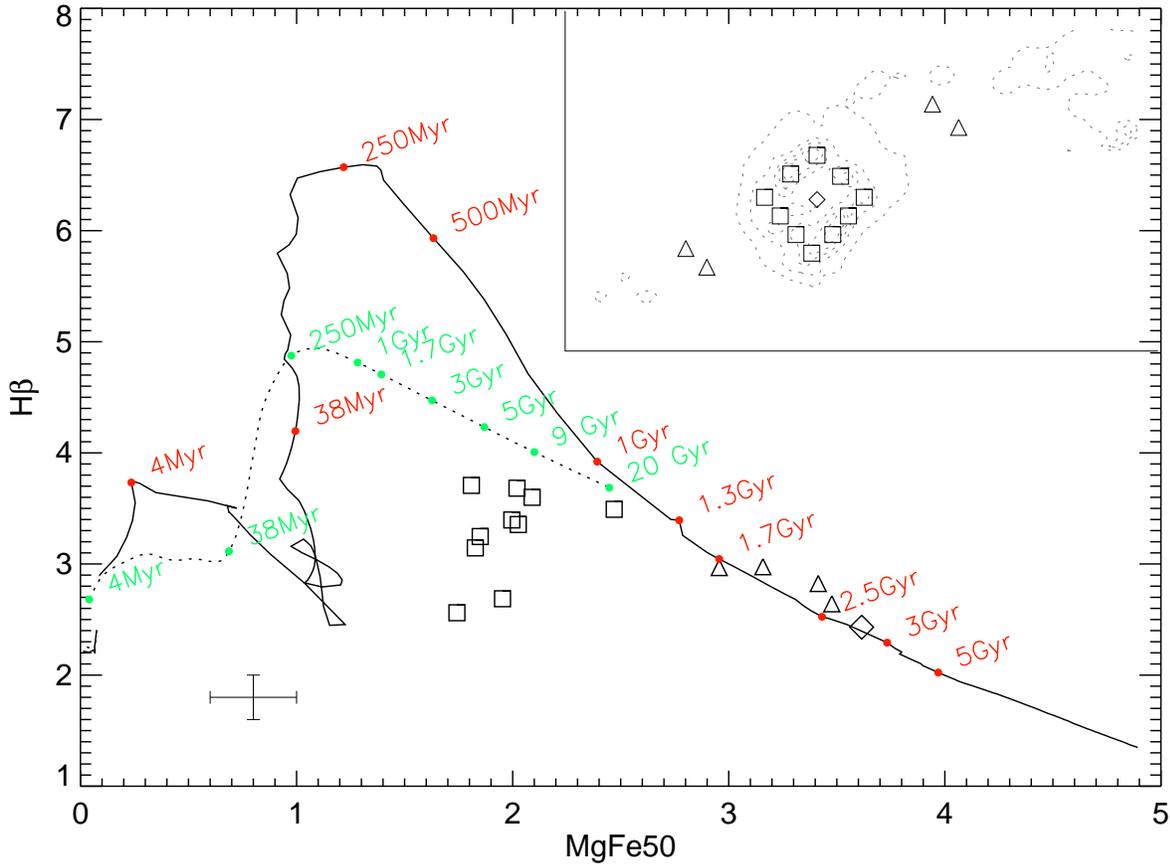}
\caption{Evolution of the H$\beta$ and MgFe50 indices with time, for
  an SSP. The inset shows the locations of the apertures taken to
  measure the data points relative to the H$\beta$ emission
  (contours). The solid line shows the track for an SSP starburst
  model, with indicative ages shown in red. The dashed line shows the track for the continuous starburst model, which has a constant star formation rate. The error bars in the bottom left are typical errors for the data points.}
\end{minipage}
\end{figure*}

\begin{figure}

\includegraphics[width=0.45\textwidth]{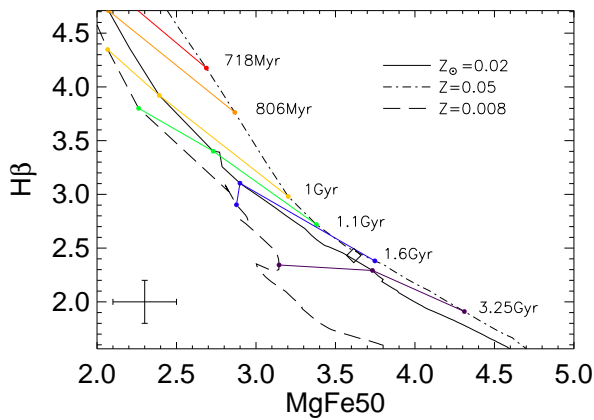}
\caption{Three SSP models of different metallicites are shown with the nucleus plotted as a diamond. The coloured lines represent lines of constant age.}
\end{figure}

\subsubsection{SSP models}
As illustrated in Fig.~12: if the H$\beta$ absorption is low, the stellar population is either old, or very young. The Mg{\emph b} and Fe5015 indices increase as a stellar population ages, so combining these indices with H$\beta$ allows a more accurate estimate of the age of a population. 
To use the model predictions of BC03 we must first make assumptions on
the star formation history of M100, for example whether the star
formation has been continuous, or has consisted of discrete
bursts. Figure~12 shows the time evolution of the MgFe50 index versus
the H$\beta$ index for a single burst stellar population (SSP; solid
line), as well as for continuous star formation (dashed line). We
chose to adopt solar metallicity throughout this paper, but Fig.~13
justifies this choice by showing the effect of sub- and super-solar
metallicity on the part of the H$\beta$ versus MgFe50 diagram relevant
to the nuclear data point. We see that it is impossible to fit this
nuclear point with a sub-solar metallicity, whereas a fit with
super-solar metallicity implies an age of less than 1.5\,Gyr. This age
is rather less attractive than the 3\,Gyr age estimated from the
curves at solar metallicity, however, we are strictly speaking unable to distinguish between the two metallicities. We have adopted to use solar metallicity in all following models, although it should be noted that increasing the metallicity to sub-solar will scale down all ages, although relative ages will not be affected.  

The basic trend in Fig.~12 is that MgFe50 increases steadily with time
for an SSP, and H$\beta$ first increases rapidly, peaks at around
250~Myr, then decreases gradually. The nuclear point (diamond) is well
fitted by this SSP at an age of 2.5-3~Gyr. At other locations this
model also appears to be a reasonable fit, such as in the bar
(triangles). The points within the ring (squares), however, are not
well fit, and so an SSP cannot be dominating the spectrum. A composite
population must be built up to match these ring points. Figure~12 also
shows the MgFe50  index versus the H$\beta$ index for a population
with a continuous star formation rate (dashed line) but this does not fit any of the points. For continuous star formation, the values of MgFe50 increase much more slowly, and values of up to 3.5 such as we observe in the nucleus and bar would not be obtainable within the lifetime of the universe.

\subsubsection{Double SSP models}

The location of the ring apertures in Fig.~12 can be explained as a
luminosity weighted combination of relatively few, but bright young
stars of a few tens of Myr in the ring, and of many more, less bright,
and much older bulge stars, a few Gyr old. To model this, we start out
by constructing rather simplistic double burst models, which we then
develop into more realistic multi-burst models in Sect.~5.3.3. There are two basic free parameters when constructing double burst models, namely the delay between the first star formation event and the second, and the relative masses (amplitudes) of the stellar populations created. Changing these two parameters affects the spectral indices in similar ways. To demonstrate this, models with the same burst amplitudes but two different delays, of 3 and 10\,Gyr, are shown in Figure~14. Increasing either the relative amplitude of the second burst or the delay between the two bursts causes the model tracks to move to the upper left in the index diagram (higher H$\beta$ and lower MgFe50). This means that we cannot fit a unique model to the nuclear ring apertures, as a number of different combinations of burst amplitudes and burst delays would fit our data points. We can, however, use such families of models to exclude certain past histories of star formation, and to identify the most plausible ones.

To start, we assume an underlying bulge and disc population in M100 as
the result of a star formation event which occurred up to 3~Gyr ago, a
timescale which coincides with our estimated age of the nuclear point
in M100. The exact time at which this initial burst of star formation
occurred is not critical, as the spectral indices change only very
gradually with time for stars over 2~Gyr. We set time, $t$, to 0 at the
end of the first star formation event. The exact nature of this
initial event (e.g., single burst, multiple burst) is also
unimportant. What \emph{is} important is that by far most of the
stellar mass was formed in this past event (see the implausibility
of the high relative second burst strength models in Fig.~14, shown in
blue). We know that star formation is occurring now in the ring from the presence of H$\beta$ emission and the low [O{\sc iii}]/H$\beta$ ratio, so for these positions in the galaxy we are justified to add the current burst of star formation, which occurs 3~Gyr after the disc and bulge population was consolidated. 

Figure~14 shows that a double burst model can reproduce the line
indices observed throughout the nuclear ring of M100 (for instance, a
model with a 3~Gyr delay and a relative second burst strength of 5 per
cent: the yellow model in Figure~14a). The location of the ring
apertures compared to the model predictions further suggests the
presence of an age gradient within the ring (see Section
6.6). However, such a model is not too attractive because it implies
that the nuclear ring has only ever had one starburst episode, which
is the one we are witnessing at present. Considering the proposed
close dynamical connection between the ring and the long-lived bar,
one would rather assume that the actively star-forming phase of the
ring had been more extended. We thus enhance our model by adding
additional bursts between the end of the past star formation event and the current burst.  

\subsubsection{Multiple SSP models}
Star formation must have stopped after the initial event to be followed only by episodic bursts, otherwise the H$\beta$ index would have become too high and the Mg{\emph b} and Fe5015 indices would have never increased to the observed values, as shown by the continuous star formation model of Figure~12. As star formation rates in nuclear rings are high, it is likely that star formation will quickly use up available gas in the ring, until more gas is supplied to the bar, causing the pattern of star formation to be episodic rather than continuous. We tested this by starting out from the model with just the current burst of star formation, and observing the behaviour of the spectral indices as extra bursts were added one at a time. The results of this can be seen in Fig.~15 for a time separation of 0.5~Gyr between bursts, and in Fig.~16 for a time separation of 0.1~Gyr. As more bursts are added, the position of the line in Fig.~15 moves vertically upwards and slightly to the left after the first extra burst, then comes back down. Regardless of the number of bursts, the spectral indices do not change significantly, as long as the time between the bursts is quite large ($\sim$0.5~Gyr). But, importantly, Fig.~16 shows what happens if the time separation is decreased to 0.1~Gyr, a number which is more plausible in a scenario of recurrent ring activity. The red line corresponds to the original model described above and shown in Fig.~14. As the number of bursts is increased, the modelled values move significantly to the left and up (higher H$\beta$ and lower MgFe50). Due to the orbital movement of stars and gas around the ring, determining the number of bursts a particular star-forming clump has experienced will be impossible. Additionally, the errors on the individual points do not allow us to be this specific about the exact number of bursts. Figure~16 shows that the ring apertures can be matched by short time separation multi-burst models. Although it appears from Fig.~16 that the variation in H$\beta$ values is due to varying numbers of bursts, the orbital movement of stars and gas around the ring will erase any signature of this in the integrated spectra. Additionally, the errors on the individual points do not allow us to be this specific about the exact index values. From the multi-burst models, it is evident that stellar age increases as one moves vertically and to the right in the index diagram, as was the case in the double burst models (see Section 6.6).

\begin{figure*}
\begin{minipage}{180mm}
\includegraphics[width=1\textwidth]{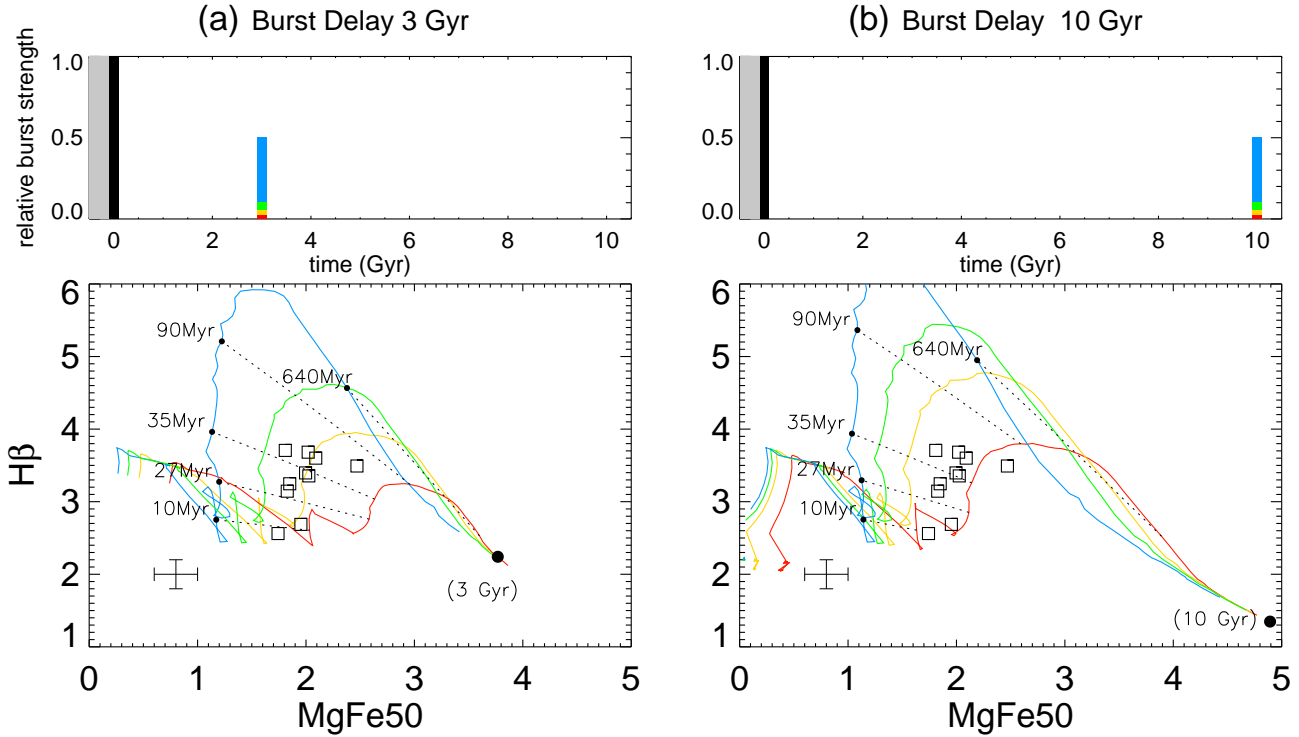}
\caption{The evolution of the MgFe50 and H$\beta$ indices with time
for four populations with varying burst strengths. The upper panels
show a representation of the star formation history for each model,
where the initial star formation event is shown by a black bar, and
other bursts are given as fractions of mass of this initial event. The
grey area represents our uncertainty at what happened before the end
of the first star formation event. The time since the first star
formation event is given on the lower axis. All bursts are
instantaneous. The delay between the initial burst and the second
burst is 3\,Gyr (a) and 10\,Gyr (b). Dotted lines of equal age are
plotted with the time labelled since 3 or 10\,Gyr (when the last burst
occurred). The age given in brackets represents the location of an SSP
of that age. The relative burst strengths for both (a) and (b) are as
follows: 0.02 (red), 0.05 (yellow), 0.1 (green) and 0.5 (blue).}
\end{minipage}
\end{figure*}

\begin{figure}
\includegraphics[width=0.45\textwidth]{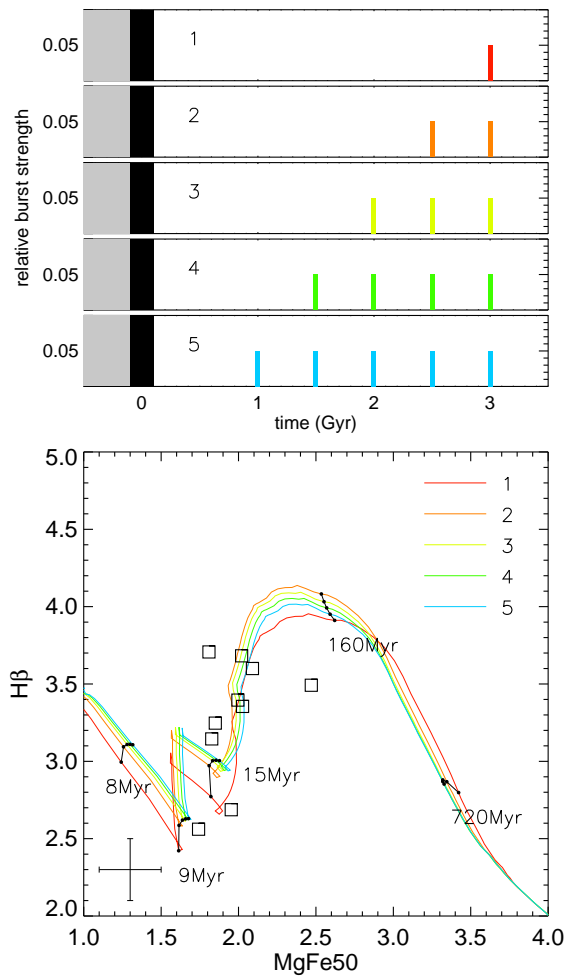}
\caption{Five different SPP models are shown, with the upper panels
representing the star formation history of the five models. Note: The
relative amplitudes of each burst are not shown to the same scale as
in Figure~14. Each coloured burst has a burst strength of 0.05
(relative to the initial burst, shown in black). The grey area
represents our uncertainty at what happened before the end of the
first star formation event. The time separation between the bursts is
0.5~Gyr. Dotted lines of equal age are plotted with the time labelled
since 3~Gyr. }
\end{figure}

\begin{figure}
\includegraphics[width=0.45\textwidth]{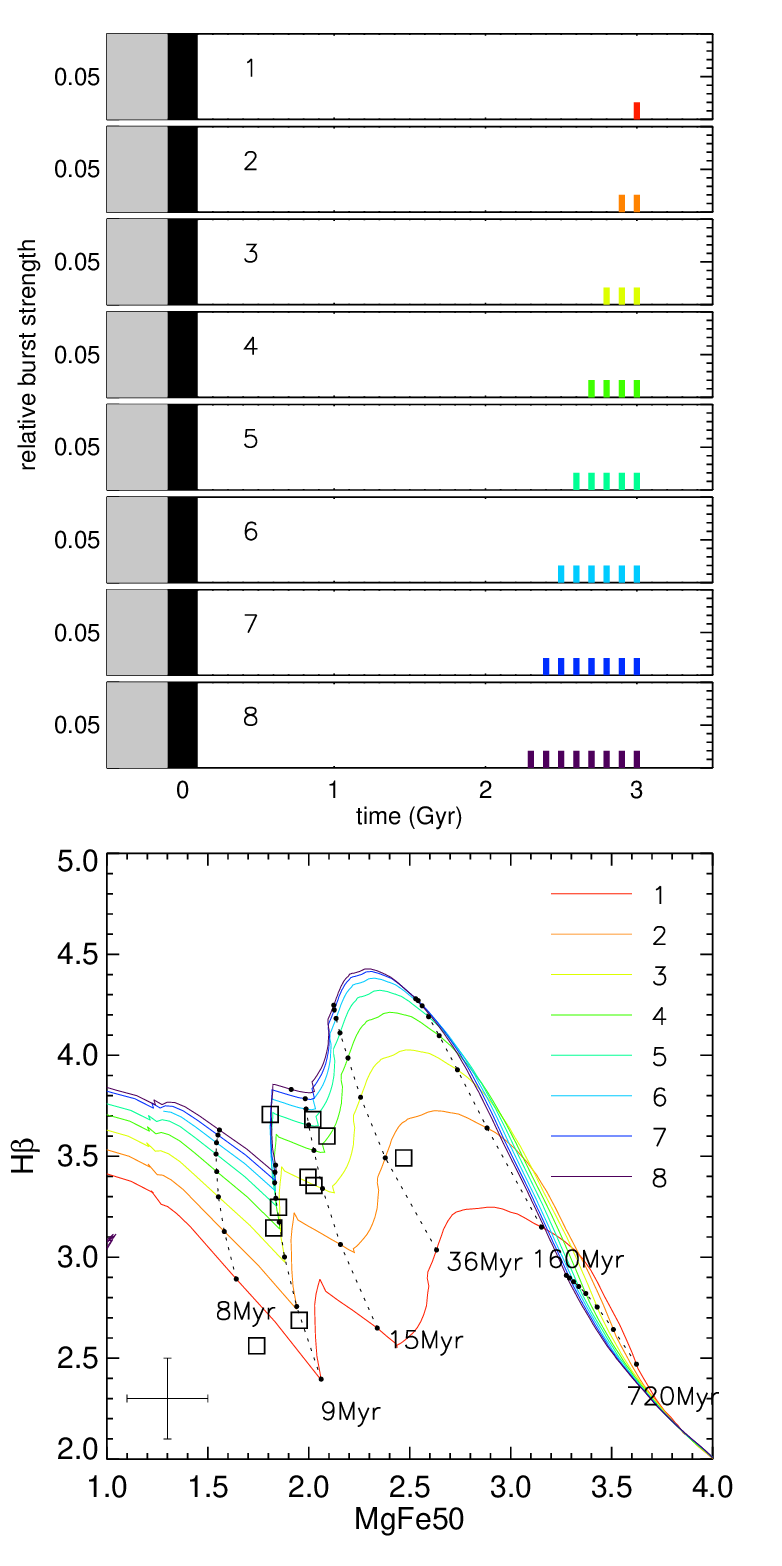}
\caption{As Fig.~15, but now eight different models are shown. The
relative amplitudes of each burst are not shown to the same scale as
in Figure~14. Each coloured burst has a burst strength of 0.02
(relative to the initial burst, which is shown in black).}
\end{figure}

\subsection{Mg{\emph b} versus Fe5015 index}
The Mg/Fe abundance ratio in galaxies is found to vary from that in
the Galactic stars normally used for calibration (Worthey, Faber \& Gonzalez 1992; Worthey 1994). There is often an over-abundance of Mg as compared to Fe in elliptical galaxies \citep{pel89,wor92,dav93}. The two elements are created predominantly in supernovae of different types: Fe in Type Ia and Mg in Type II supernovae. Type Ia supernovae are due to the accretion of a binary companion onto a white dwarf - a relatively long process. Type II supernovae are produced from the core collapse of massive stars which have much shorter lifetimes. An overabundance of Mg suggests that the population is dominated by massive stars, which can be the result of a flat IMF, short timescales for star formation, or a low fraction of binaries experiencing Type Ia supernovae \citep{wor92}. 

Figure~17 shows the evolutionary track for the Mg{\emph b} and Fe5015
indices for an SSP model, with the data points from the centre and the
bar overplotted. The points are not well fitted by the model, and have
larger Mg{\emph b} and/or smaller Fe5015 values, indicating that the
abundance ratio of M100 is different from that in the local stars used
in the models. Combining the indices as in the previous section helps
to remove this problem when determining ages.\footnote{\citet{thom06}
  found that H$\beta$ was the least abundance-sensitive Balmer line
  age indicator.} In Fig.~18 we plot the indices for apertures on the
ring compared to three multi-burst models. These points fit much
better, considering that the exact model parameters are uncertain.

The stars in the centre and along the bar all have Mg/Fe ratios which
are too high, suggesting that the initial star formation event in the
galaxy ($\sim$3~Gyr ago) occurred over short timescales, and most of
the stars were formed before enough Type 1a supernovae had enriched
the ISM with Fe, which can take several Gyr \citep{wor92}. The new
stars forming in the nuclear ring do not have an over-abundance of Mg,
which makes sense as enough Type 1a supernovae from the original burst
should have occurred by now. This offers support for the assumption
that the gas creating these new stars has been brought in from the
disc under the influence of the bar. \citet{beau97} studied the
abundance gradient Mg/Fe in the disc of M100 and found that it drops
at larger radii. Peletier et al. (in preparation) show from SAURON
data that the Mg/Fe in early-type spirals is generally larger than
solar, with considerable scatter, however. The scenario they suggest,
with quick bulge-formation and subsequent additional bursts, agrees
with our model for  M100.

Thomas, Maraston \& Bender (2003) and Thomas, Maraston \& Korn (2004)
have provided SSP models (which we label TM04) with variable abundance ratios, which we have used to plot our Mg{\emph b} against Fe5015 measurements compared to three models with different $\alpha$/Fe\footnote{Mg is taken to be representative of all $\alpha$ elements.} ratios for solar metallicity (Figure~19). As the $\alpha$/Fe ratio increases, the Mg{\emph b} against Fe5015 track moves down and to the right, encompassing the data points from the bar and nucleus. These results confirm the discussion in the previous paragraph, whereas the ages derived from these particular models are consistent with those determined in Section 5.3.1. 
\begin{figure}
\includegraphics[width=0.45\textwidth]{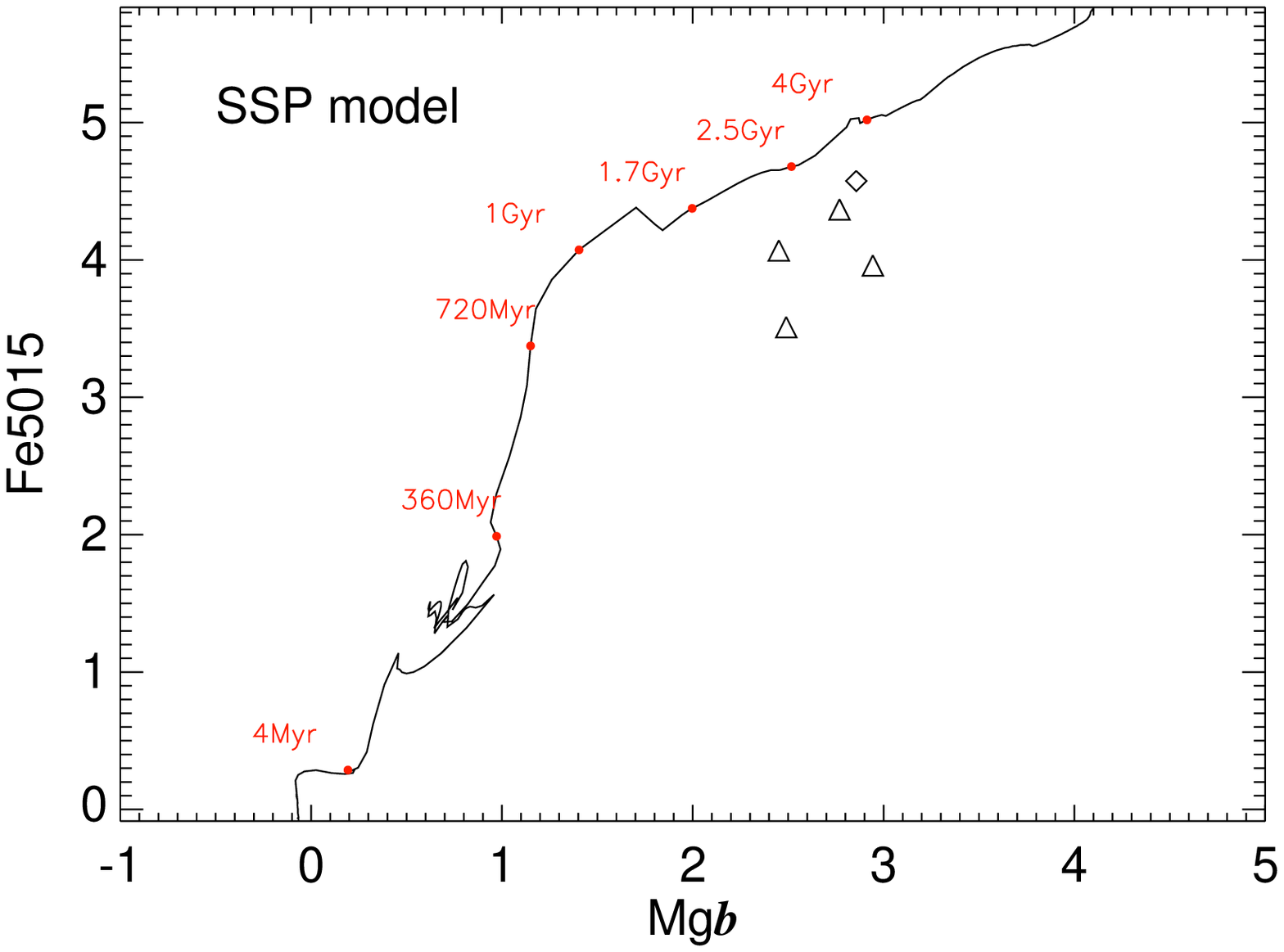}
\caption{The evolution of the Mg{\emph b} and Fe5015 indices with time for a SSP, with the points from the centre and bar plotted.}
\end{figure}

\begin{figure}
\includegraphics[width=0.45\textwidth]{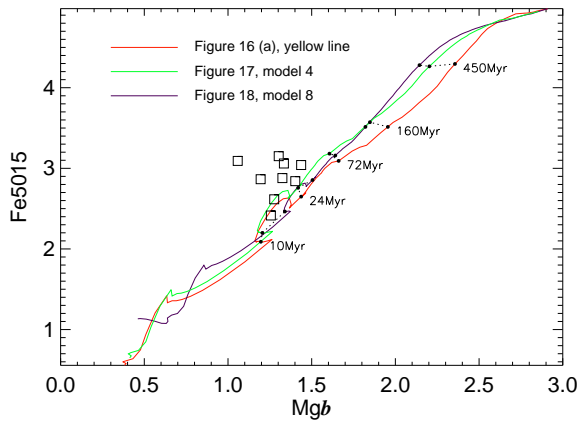}
\caption{The evolution of the Mg{\emph b} and Fe5015 indices with time for three models, with the points within the ring plotted. Dotted lines of equal age are plotted with the time since 3~Gyr (when the last burst occurred) labelled. }
\end{figure}

\begin{figure}
\includegraphics[width=0.45\textwidth]{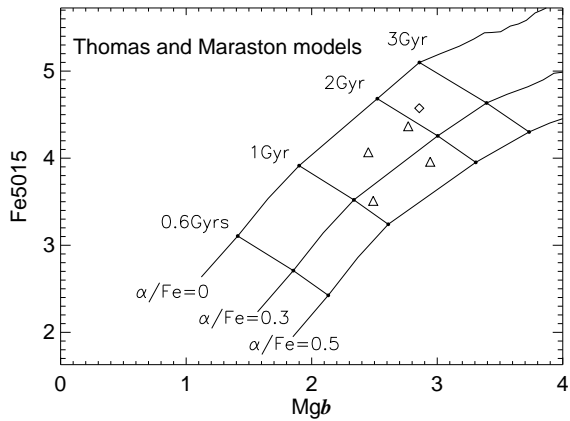}
\caption{The evolution of the Mg{\emph b} and Fe5015 indices with time for a SSP, using the TM04 models with varying $\alpha$/Fe ratios.}
\end{figure}

\section{Discussion}
\subsection{Orbits in the inner region}
                                                                                              
The twisting of the velocity contours of the gas, previously observed
and now confirmed, has been explained by K00 as a combination of
streaming motions caused by the inner part of the bar and by an
associated spiral
density wave within the central region. The stellar kinematics, which,
as described in more detail above (Section 4.2), follow the general
characteristics of the gaseous velocity field, demonstrate that the
stars are also undergoing streaming motions due to these two separate
effects. According to non-linear theory, stars belonging to the bar follow highly elongated orbits aligned parallel to the bar, which leads to
the observed non-circular motions as we see at a radial distance of
$\sim5$ arcsec from the nucleus. This had been discussed for the gas
in the inner part of the bar in M100 by K00, but is now reported for
the first time in the stars. Knapen et al. (1993) had earlier found the same effect on the gas in the large
bar of M100, but the detailed study of its stellar counterpart is
beyond the quality  of
our data set.

The second set of velocity contour twists, at radii of $\sim12$ arcsec, was explained by K00 as streaming motions in the gas due to the effects of a spiral density wave, which lead to a gradual
skewing of the gas orbits. We confirm these motions in the gas, but
also observe them clearly in the stellar velocity field. In fact,
their radial position coincides with the region where our kinemetric
analysis of the gas velocity field suggests a localised radial region
of inflow - presumably due to material which flows through the spiral
arm wave and in doing so moves to smaller radii. We note that K00 made
a qualitative comparison of the observed and modelled gas velocity
fields and found that the streaming features due to both bar and
spiral armlets were reproduced well in the model.

\subsection{Age distribution of stars around the ring}
One similarity between various models of ring formation is the net
inflow of gas into the ring near the contact points between the ring
and the dust lanes. To investigate whether this inflow directly
influences the massive star formation, we can measure ages for the
brightest clumps in a nuclear ring to see whether there are any
underlying azimuthal trends. Ryder, Knapen \& Takamiya (2001) carried out such a
study of the nuclear ring in M100, using UKIRT NIR spectroscopy, and
discovered marginal evidence for a gradient in clump ages around the
ring, with the youngest clumps found at the contact points. This would
support a picture where the gas inflow near the contact points
immediately causes gravitational collapse and massive star formation,
rather than one in which the gas inflow gradually fills the ring with
gas, and where massive star formation would then occur more
stochastically around the ring. Mazzuca et al. (2006) investigate this
across a sample of 22 nuclear rings and find that most follow the
latter scenario, with a small minority of the rings showing evidence for significant azimuthal age gradients. 

Our data allow an analysis of such gradients in the ring of M100 using
two different techniques, a comparison of the stellar absorption line
strengths with the BC03 models, and a measurement of the H$\beta$
emission equivalent width (EW). The EW of the H$\beta$ emission of an
SSP decreases extremely quickly and monotonically with time in the age
range 2 to 6~Myr (e.g. Leitherer et al. 1999; Terlevich et
al. 2004). Because this basic behaviour is independent of metallicity
and of the shape of the IMF, we can use the EW as a sound relative age
indicator, and do so for the emitting regions around the ring.  We
determine the H$\beta$ EW by measuring the flux of the H$\beta$
emission line, after correcting for the velocity dispersion and
instrumental resolution. To estimate the continuum flux, we determine
the median flux level across the complete spectral range from the
emission corrected stellar spectra. Dividing the flux of the line by
this continuum level gives the EW of the line. Figure~20 shows the
H$\beta$ EW map of the nuclear ring region, whereas Fig.~21 shows the
EWs for the nuclear ring apertures defined in Fig.~22, plotted as a
function of their PA, in the same format as used by Mazzuca et
al. (2006) for their sample of 22 galaxies. 

Figure~21 indicates that the highest EWs, thus youngest ages, occur near two positions along the ring, $\sim 180^\circ$ apart. EWs between these two positions are lower, indicating older ages, while the Western apertures, between PA$=170^\circ$ and $300^\circ$, show a monotonically decreasing set of EW values. 

Figure~22 plots, for the same apertures within the ring, the H$\beta$ against MgFe50 absorption line strengths. Although it is impossible to give exact ages for each clump due to the uncertainties in the star formation history, a vertical sequence in colours is prominent, with red and orange points at the bottom of the plot, and green and blue colours towards the top. Relating to the ages given by the model, this sequence is readily interpreted as the same bipolar age sequence as seen in EW (younger to older from bottom to top). The red and orange points indicate an age of around 10-30~Myr, and the green and blue points around 30-50~Myr. The absolute ages of these stars are of course model dependent, but the differences between ages are significant.

Using two independent methods we have therefore found convincing and
consistent evidence for the presence of a bipolar age gradient in the
nuclear ring of M100, with the youngest stars located near the contact
points, where the gas flowing from the disc and through the bar (Paper
I) reaches the ring. Massive star formation occurs here, and as these
newly formed H{\sc ii} regions move along the ring, star get older and gas emission faints. The age gradient along the ring we inferred using the
EW of the H$\beta$ emission and the stellar absorption line strengths is
consistent with the time needed for gas and star to move from one
contact point to the other along the ring, which is a few tens of Myr
(assuming a ring radius of $\sim5$\,arcsec $=350$\,pc and a velocity of
$\sim 100$\,km\,s$^{-1}$).

\begin{figure}
\includegraphics[width=0.3\textwidth]{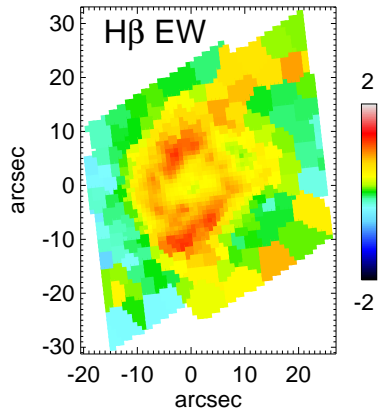}
\caption{The H$\beta$ emission equivalent width map for the ring region. The colour scale is logarithmic.}
\end{figure}
\begin{figure}
\includegraphics[width=0.45\textwidth]{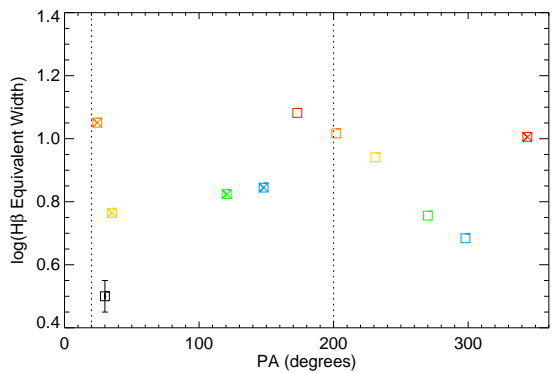}
\caption{The H$\beta$ emission equivalent widths of points along the ring, plotted against their PA. The contact points are marked with dashed lines. Figure~22 gives the location of these points relative to the ring.}
\end{figure}

\begin{figure}
\includegraphics[width=0.5\textwidth]{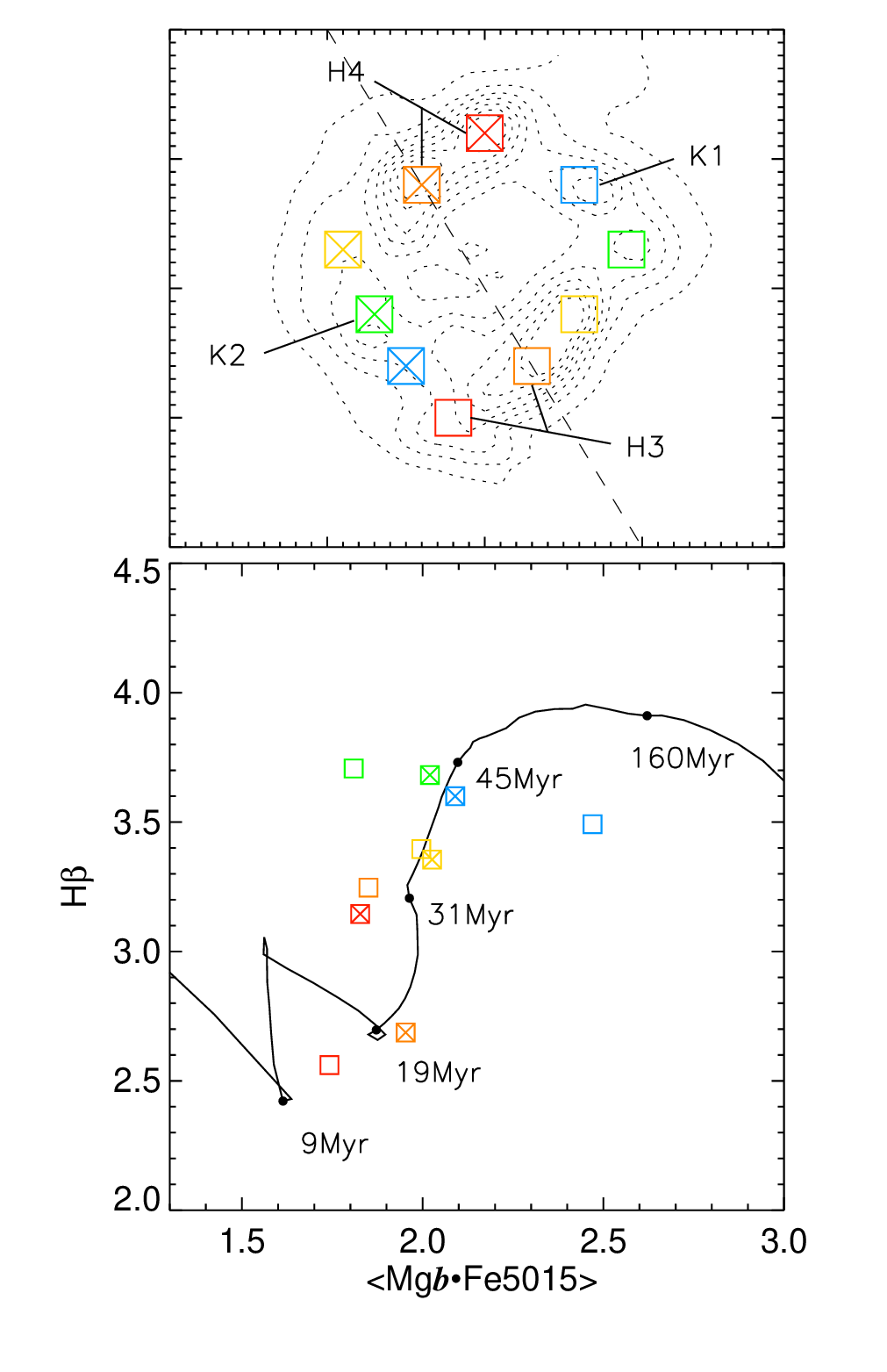}
\caption{Upper Panel: the location of the points of Fig.~21 within the
  ring, shown relative to H$\beta$ emission in the central region. K1,
  K1, H1 and H2 refer to the locations of regions K1, K2, H$\alpha$1 and
  H$\alpha$2 as given by K95b. Lower panel: the position of these points in
  the index diagram, plotted with a double burst model for
  comparison. }
\end{figure}

\subsection{The star formation history of the bar and nuclear ring}

From the comparison of our SAURON observations with the
theoretically
 expected absorption and emission line behaviour of
evolving stellar
 populations in the central region of M100, we can
construct a coherent
 picture of its star formation history. The
different parts of this
 comparison have been described in detail in
the preceding sections, but
 we aim to summarise the results here and
place them into the context of
 our understanding of the bar and the
nuclear ring which it fuels.
 The main premise of this understanding
is that the asymmetric potential
 outlined by the bar extracts
angular momentum from the gas, which flows
 inwards as a result
(Prendergast 1983; Athanassoula 1992; Shlosman, Begelman \& Frank 1990). The
inflow of gas slows down in the vicinity of a pair of
 ILRs, and
between them, near the inner ILR, massive star formation
 occurs as a
result of the gravitational collapse of accumulated gas, and
 does so
in a ring-like region which we observe as the nuclear ring.
 This
picture has been confirmed from the observed morphology at a
 variety
of wavelengths, from kinematics in the different emission lines,
 and
from numerical modelling (K95a,b; K00). The basic dynamics of a
bar-induced nuclear ring (cf. Athanassoula 1992; K95a; Heller \&
Shlosman 1996), as well as the resulting massive star
formation properties, have been observed and modelled for a variety
of
 other galaxies (e.g., Benedict, Smith \& Kenney 1996; Regan,
Vogel \& Teuben 1997; Jogee et al. 2002a,b; Maciejewski 2004a,b), and
can be considered a generally valid
 interpretative framework for
such systems.

In the first publication resulting from our SAURON data set of M100
(Paper~I) we in fact corroborated this framework further by showing
graphically how relatively cool gas flows in through the bar, and into
the nuclear ring. The locations of the cool gas coincide exactly with
those of the current massive star formation as observed through its
H$\beta$ emission, but are significantly offset in the bar region from
the curved dust lanes which trace the location of the shocking and
compression of the gas.

The current paper attempts to use both new kinematic and morphological
measurements, and the strengths of emission
and absorption line features due to stars and gas, to further elucidate
the history of star formation within the bar and nuclear ring region in
M100. The absorption line indices of H$\beta$ and MgFe50, first of
all, show how the light from the nucleus of M100 and from its bar can be
described simply by an old stellar population, whose age we estimate to
be around 3~Gyr (Figure~12). We thus confirm that the bar and the nucleus
(bulge) are formed by old stars, as expected, and as confirmed in, e.g.,
the {\it SST} 3.6 and 4.5\,$\mu$\,m maps (Figure~3). The light from
the nuclear ring, however, cannot be described by any single past
star formation event, nor by continuous star formation (Figure~12).
For the ring, we thus develop alternative models, and find that the only
way to reproduce the observed intermediate values of MgFe50 in
combination with low H$\beta$ is to introduce a past star formation
event which stopped around 3~Gyr ago (where this figure is given by the
age of the stars in the bar and nucleus). We cannot constrain the form
of that event, whether it was one or several discreet bursts or a more
continuous star formation, because each of these scenarios would by now
have led to an identical old population, but the line index measurements
in the ring do dictate an end to this episode of disc and bulge
formation roughly at the time stated above. In addition, our model
invokes recent and current star formation in the ring, which we of course
know must exist to give the observed massive star formation rate. We can
reproduce the observed line indices by a combination of the past
event and one current burst of star formation (Fig.~14), but this
particular model has the conceptual disadvantage that the ring would
have been quiescent from 3~Gyr ago to very recently, which is at odds
with the scenario in which the (long-lived and old) bar continuously
feeds the ring with gas.

Our preferred model, then, reproduces the observed line indices by
combining our past  star formation event with a number (around five, say)
of discreet but rapidly repetitive bursts, separated by, say,
100~Myr. In addition, our line indices and, independently, H$\beta$
emission EW measurements show how the ages of the individual
star-forming regions vary systematically azimuthally around the ring,
with the youngest regions lying near the contact points where the dust
lanes connect the bar and ring (discussed in more detail in
Sect.~6.2). Given the nature of our data and the time-behaviour of
line indices in such complex populations, we cannot constrain the
parameters describing this series of bursts to high precision: for
instance, five bursts of a certain intensity give the same line indices
as six with slightly lower intensity. But the model we present here is
entirely plausible, because it represents the star-forming nuclear ring
as an entity that is regularly undergoing massive starbursts, and has
done so during a period of time of order $5\times10^8$\,yr. This
timescale connects in a consistent manner the presence of a bar, which
is presumably stable and long-lived (e.g., Jogee et al. 2004), and the
nuclear ring, whose timescales for both formation and, once formed,
stability are entirely consistent with the modelling performed by
K95a. Linear instability analysis (Elmegreen 1994) also suggests a
timescale of $\sim$10$^8$~yrs for the lifetime of star formation in a nuclear ring. We thus confirm that nuclear rings are a stable configuration in
which a bar feeds the ring with relatively cool gas flowing in from the
disc, and in which the ring itself can host a number of consecutive
massive star formation events.

\subsection{Inner part of the bar}

\begin{figure}
\includegraphics[width=0.45\textwidth]{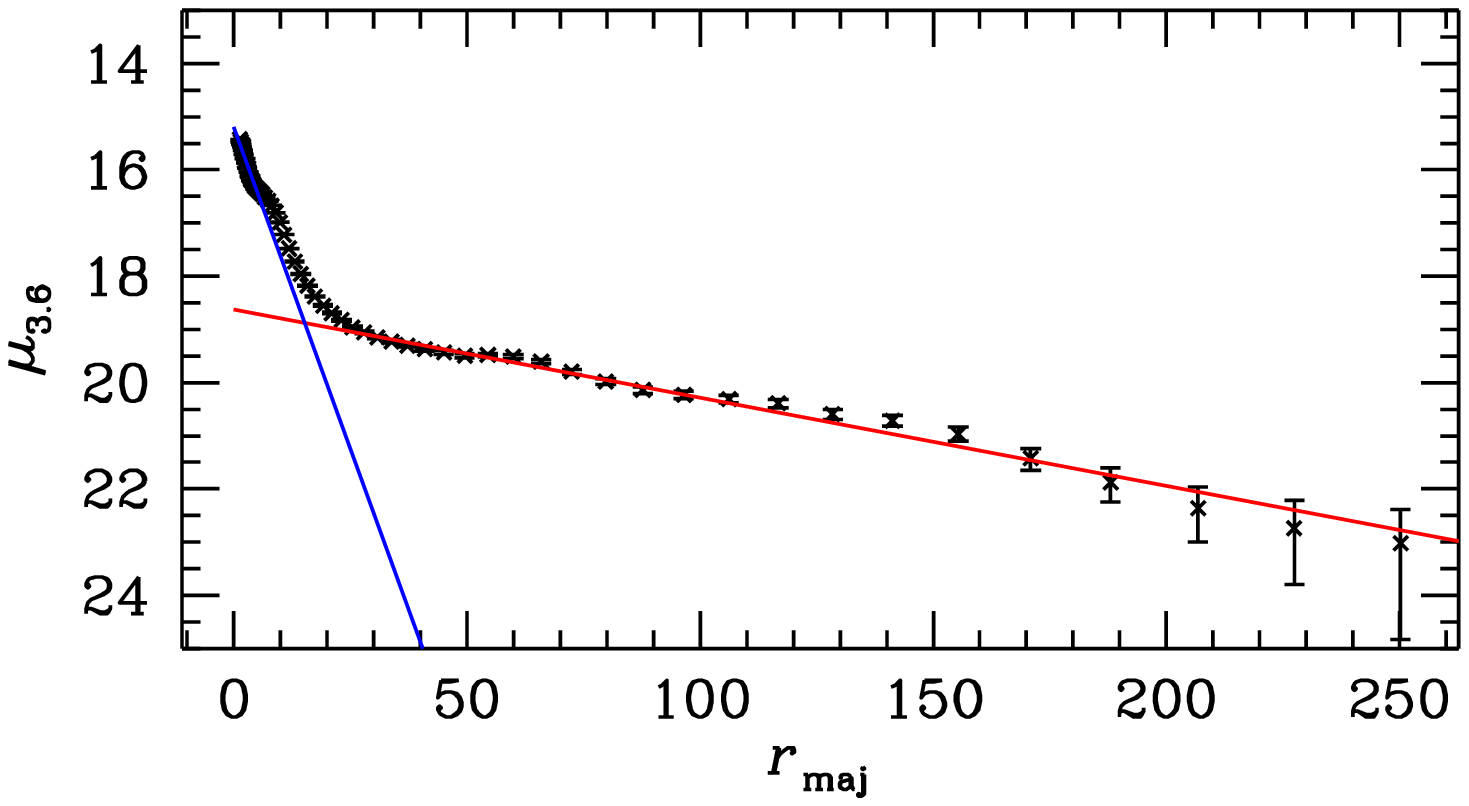}
\includegraphics[width=0.45\textwidth]{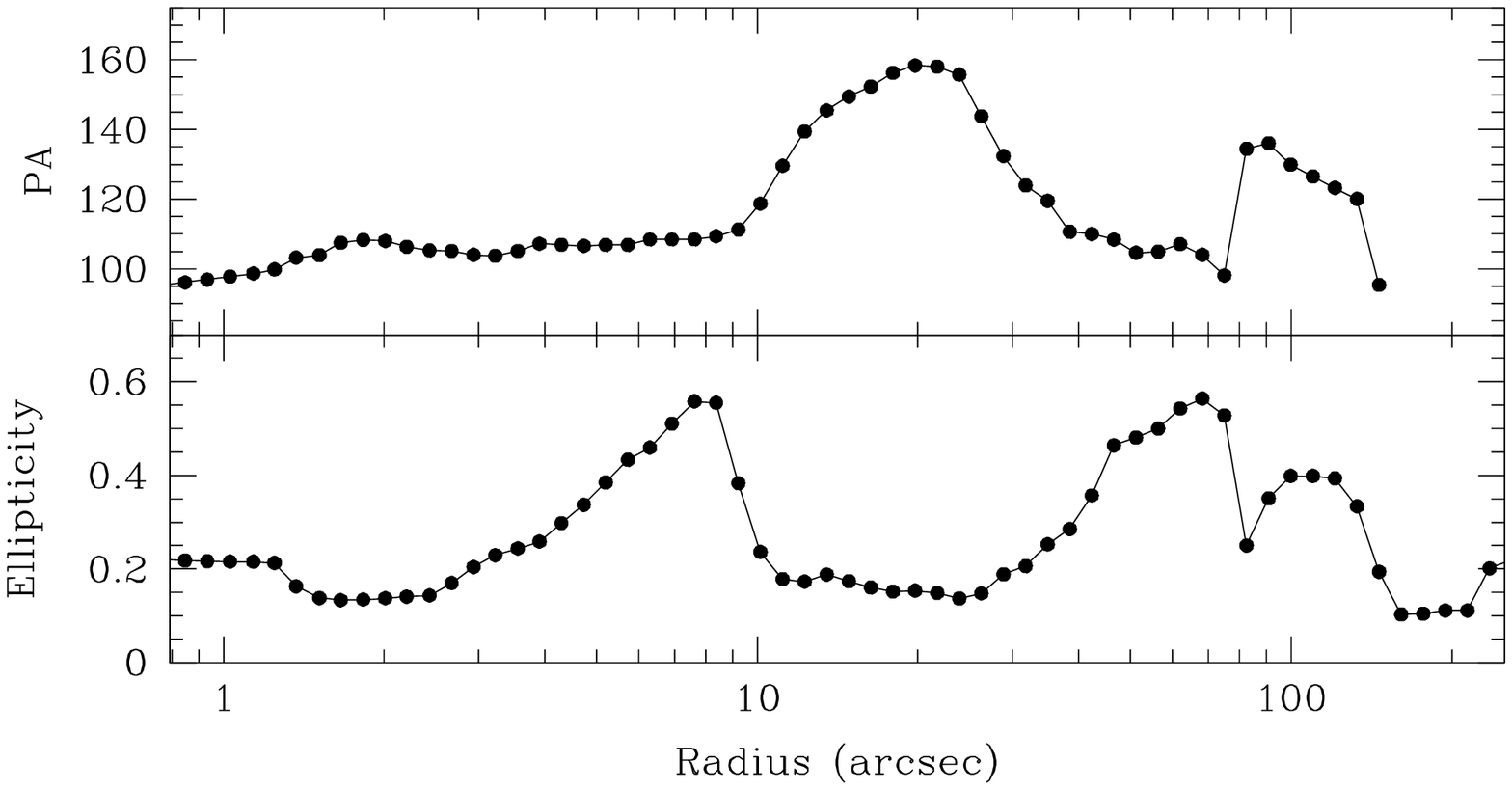}
\caption{Radial profiles of surface brightness $\mu$ ({\it top}, with
  linear radial scale), and PA and ellipticity ({\it middle} and {\it
  bottom}, with logarithmic radial scale) as derived from ellipse
  fitting to the {\it SST} 3.6\,$\mu$m image of the complete disc of
  M100. The $\mu$ profile has been calculated with fixed PA=30$^\circ$
  and $\epsilon=0.149$. Exponential scale length fits are indicated
  for the bulge (blue) and disc (red), where the latter was fitted
  between radii of 35 and 150\,arcsec. The two parts of the bar can be
  clearly seen as distinct rises and falls in the ellipticity
  profile.}
\end{figure}

The 3.6 and 4.5 \,$\mu$m IRAC images allow us to revisit the parameters
of the inner and outer parts of the bar of M100, which were inferred
by K95a to be two parts of the same dynamical structure (i.e., bar). This
conclusion was reached originally because the two bars have the same PA and a very similar radial ellipticity
behaviour, and on the basis of the overall morphology of stars and dust lanes. The morphology of the inner part of the bar is severely
distorted by the combined influence of dust extinction and emission by
young stars, and this is even the case at 2.2\,$\mu$m (see K95a for a
detailed discussion). We can now use the IRAC images to revisit this
issue, and for this purpose have run the ellipse fitting task in IRAF's
STSDAS on the 3.6\,$\mu$m image. We were able to fit the radial runs of
the surface brightness $\mu$, the PA of the major
axis, and the ellipticity across the complete range in radius from the
inner resolution limit near 2~arcsec to well into the optical disc,
although the fits are less well constrained at the outer limits. The
resulting fits, shown in Fig.~23, clearly confirm the
results obtained in K95a on the basis of a $K$-band image of the inner
part of the bar and an $I$-band image of the disc. The main
characteristics are the two parts of the bar, defined by a very clear
rise and fall of the ellipticity, in both cases from the disc value of
0.14 (corresponding to the inclination angle of the disc of 30$^{\circ}$) to a maximum value of $\epsilon=0.56\pm0.015$, which occurs
at radii of $r=6.5$ and $r=55$\,arcsec for the inner and outer
parts of the bar, respectively. The PA is stable and identical (at a
value of PA=107$\pm$3\,degrees) at the location of both these peaks in
ellipticity, but is higher in the region between them. 

The radial
 surface brightness profile shows the locations of
 both
parts of the bar, but a detailed discussion of these and other
features of this profile is outside the scope of this paper. We
performed a bulge/disc decomposition on the 3.6\,$\mu$m surface
brightness profile, the results of which are shown in Fig.~23. The
bulge fit yields an exponential scale length of 4.5\,arcsec, whereas
the disc yields 65.6\,arcsec. This latter value is not very different
from the values given by Beckman et al. (1996), who report scale
lengths between 67.5 and 76.4\,arcsec for different optical wavebands
(albeit determined over a larger radial than our 3.6\,$\mu$m
value). The fact that the scale lengths do not vary much, even going
out to 3.6\,$\mu$m, reinforces the conclusion by Beckman et al. (1996)
that the galaxy does not contain much dust, except in localized areas
(spiral arms).

In
summary, we confirm the conclusion from K95a, but now on the basis of
much better observational material, that the inner and outer parts of
the bar of M100, located inside and outside the star-forming nuclear
ring, have identical (within the error margins) radial elliptical and
PA profiles.
The inner part of the bar is not visible in the SAURON morphology
maps, as it is primarily composed of older stars (K95a). The influence of the bar can be seen in the stellar kinematics, as a small deviation near the minor axis. The stellar kinematics are derived from the stellar absorption features which contain a contribution from these older stars.  

\subsection{Nuclear ring morphology}
The morphology of the nuclear ring changes dramatically with
wavelength in the IRAC and 24\,\,$\mu$m MIPS images considered
here. Whereas the inner part of the bar and the emission regions K1
and K2 at its ends dominate the 3.6\,\,$\mu$m image, the star-forming
ring becomes gradually more prominent with increasing wavelength, as
seen clearly in the colour index maps (Fig.~4) where the ring is
consistently red. Because the emission from the nucleus is diminishing
with increasing wavelength, the 8\,\,$\mu$m image almost exclusively
shows the ring, in which K1 and K2 are especially prominent. We
interpret the tadpole shape of these regions as due to the emission
from the well-known miniature spiral arm fragments due North and South
of the nucleus, and labelled H$\alpha$3 and H$\alpha$4 in K95a (see Fig.~22).
Comparison with the 24\,\,$\mu$m image, at much lower spatial
resolution, shows that the Northern spiral armlet, H$\alpha$4, is
a much more active star-forming region than its Southern counterpart
H$\alpha$3. This is consistent with, but could not reliably have been deduced
from, the other data: for instance, K95b found that H$\alpha$4 was not only a
bit brighter, but also somewhat dustier than Ha3, and the 8\,\,$\mu$m
image also shows more emission from H$\alpha$4 than H$\alpha$3.

\subsection{K1 and K2}
                                                                                
The emission from the nucleus dominates at the shorter IRAC
wavelengths, and emission from the two symmetrically placed regions at
the ends of the inner part of the bar is equally strong in the Western
(called K1 in K95a) and the Eastern (K2) sites. These were described
by K95a to be symmetric regions of star formation, where K2 was
postulated to be more obscured by dust than K1. The ages of K1 and K2
are similar, between 15 and 25~Myr, as modelled on the basis of $UBV$
imaging (K95a) and near-IR spectroscopy (Ryder \& Knapen 1999). The ages derived from the absorption line indices from the SAURON data are model dependent, but K1 and K2 are found to have similar H$\beta$ absorption indices (indicating similar ages).
                                                                                
The {\it SST} imaging confirms that K1 and K2 are intrinsically very
similar. By measuring the flux within two matched apertures of 2.4
arcsec in diameter centred on K1 and K2, we find that indeed the flux
ratio K2/K1 is close to unity. The ratios (K2/K1 flux) measured in
this way are 0.88 for H$\beta$ (using the SAURON data), 1.13 for
H$\alpha$, 0.95 at $K$ (using images from K95a for H$\alpha$ and $K$),
0.97 at 3.6\,\,$\mu$m, 1.05 at 4.5 and 5.8, 1.09 at 8.0, and finally
1.29 at 24\,\,$\mu$m. These ratios are consistent with K2 being more
dusty than K1, and in fact the ratios at H$\alpha$ and H$\beta$ allow
an estimate of dust extinction a factor of 1.2  higher  at K2 than at K1. K2
is forming more stars than K1, as indicated by the most direct SF
tracer, the 24\,\,$\mu$m image (Calzetti et al. 2005), and as
corroborated by the high H$\alpha$ K2/K1 ratio. The slow increase in
the K2/K1 flux ratio from 2.2 to 8\,\,$\mu$m is consistent with this
picture. We should point out that we did not attempt to subtract the
underlying emission from disc, bulge, and nuclear ring from the fluxes
at K2 and K1. 

\section{Conclusions}
This paper presents SAURON integral field observations of the central
region and bar of M100, which yields the gaseous and stellar
kinematics, and emission and absorption line strength maps.  We
complement this data set with {\it SST} near- and mid-IR imaging of
M100. The basic
emission line morphology is that of the well-known nuclear ring which
shines brightly in H$\beta$, and which is linked to H{\sc ii} regions
at the Eastern end of the bar by a thin arc of ionised gas, running
alongside but offset from the dust lanes (see also Paper~I). In [O{\sc
iii}] the morphology shows a central peak, while generally tracing the
H$\beta$ emission.

The H$\beta$ and [O{\sc iii}] gas kinematics show regular rotation
along the bar, equivalent to an inclined rotating disc, but with
characteristic kinks in the velocity field along the bar minor axis,
near the radius of the nuclear ring.  The stellar velocity field is
similar to that of the gas in the bar, and the non-circular
motions are only slightly less evident near the centre, and
practically as strong as in the gas along the bar minor axis.
As discussed in more detail in Paper~I, the H$\beta$ gas velocity
dispersion shows a lower value where the star formation occurs, in the
ring and at the ends of the bar, and also alongside the Eastern
dust lane. The [O{\sc iii}] gas velocity dispersion shows a slight
decrease in the ring, and has a much higher central value than
H$\beta$. 

The Mg{\emph b} and Fe5015 absorption line indices show lower values
within the ring suggesting that a younger stellar population is diluting
the deep absorption features of the older bulge population. The
H$\beta$ absorption line index shows a broken ring, with the breaks
occurring where the H$\beta$ emission is the strongest. All three
absorption line indices, as well as  the H$\beta$ EW,
show  that the youngest stellar populations are found at the
contact points between the ring and the dust lanes.

Detailed modelling of the line strengths shows that old stars are
present in addition to the young population which dominates the
appearance of the ring, mostly so in emission lines like that of
H$\beta$. These old stars must have been formed in a past
star formation event which produced the bulk of the mass, and which
stopped some 3~Gyr ago, a constraint set by the age of the stars in
the bar and the nucleus. Our best-fitting model is one in which the
current star formation is but the latest of a series of relatively
short bursts of star formation which have occurred for the last
500~Myr or so.  A clear bi-polar azimuthal age gradient is seen within the ring, with the youngest stars
occurring near where the bar dust lanes connect with the ring.

Our kinematic and morphological results thus confirm a picture in
which the nuclear ring in M100, considered typical, is fed by gas
flowing in from the disc under the action of the bar, is slowed down
near a pair of resonances, and forms significant amounts of massive
stars. Detailed stellar population modelling shows how the underlying
bulge and disc were put in place a number of Gyr ago, and that the
nuclear ring has been forming stars since about 500~Myr ago in a
stable succession of bursts. This confirms that nuclear rings can form
under the influence of a resonant structure set up by a bar, and
proves that they are stable features of a galaxy rather than one-off
starburst events.

\section{Acknowledgements}
We thank Isaac Shlosman and Tim de Zeeuw for comments on an earlier version of the
manuscript.  Based on observations made with the WHT, operated on the
island of La Palma by the Isaac Newton Group in the Spanish
Observatorio del Roque de los Muchachos of the Instituto de
Astrof\'{i}sica de Canarias, and with the {\it SST}, which is operated
by the Jet Propulsion Laboratory, California Institute of Technology
under a contract with NASA. We thank the SAURON team for making the
instrument and the associated software available for this
collaborative project. SAURON was made possible through grants from
the Netherlands Organization for Scientific Research, the Institut
National des Sciences de l'Universe, the Universities of Lyon~I,
Durham, Leiden, and Oxford, the British Council, the Particle Physics
and Astronomy Research Council, and the Netherlands Research School
for Astronomy NOVA. ELA is supported by a PPARC studentship, and thanks the Kapteyn Astronomical Institute for its hospitality during a few visits. JHK acknowledges the Leverhulme Trust for the award of a Leverhulme
Research Fellowship.

\label{lastpage}


\begin{thebibliography}{99}
\bibitem[\protect\citeauthoryear{Allard et al.}{2005}]{allard} Allard, E. L., Peleter, R. F., Knapen, J. H. 2005, ApJ, 633, L25 (Paper~I)
\bibitem[\protect\citeauthoryear{Athanassoula}{1992}]{ath92} Athanasoula, E. 1992, MNRAS, 259, 345
\bibitem[\protect\citeauthoryear{Bacon et al.}{2001}]{bac01} Bacon, R., et al. 2001, MNRAS, 326, 23
\bibitem[\protect\citeauthoryear{Barth, Ho \& Sargent}{Barth et al.}{2002}]{bar02} Barth, A. J., Ho, L. C., Sargent, W. L. W. 2002, AJ, 124, 2607 
\bibitem[\protect\citeauthoryear{Batcheldor et al.}{2005}]{bat05} Batcheldor, D., et al. 2005, ApJS, 160, 76
\bibitem[\protect\citeauthoryear{Beauchamp \& Hardy}{1997}]{beau97} Beauchamp, D., Hardy, E. 1997, AJ, 113, 5
\bibitem[Beckman et al.(1996)]{1996ApJ...467..175B} Beckman, J.~E., 
Peletier, R.~F., Knapen, J.~H., Corradi, R.~L.~M., Gentet, L.~J.\ 1996, 
ApJ, 467, 175 
\bibitem[\protect\citeauthoryear{Benedict, Smith \& Kenney}{1996}]{bene96} Benedict, G. F., Smith, B. J., Kenney, J. D. P. 1996, AJ, 111, 1861
\bibitem[\protect\citeauthoryear{Bruzual \& Charlot}{2003}]{bc03} Bruzual, G., Charlot, S. 2003, MNRAS, 344, 1000 (BC03)
\bibitem[Calzetti et al.(2005)]{2005ApJ...633..871C} Calzetti, D., et al.\
2005, ApJ, 633, 871
\bibitem[\protect\citeauthoryear{Cappellari \& Copin}{2003}]{cap03} Cappellari, M., Copin, Y. 2003, MNRAS, 342, 345
\bibitem[\protect\citeauthoryear{Cappellari \& Emsellem}{2004}]{cap04} Cappellari, M., Emsellem, E. 2004, PASP, 116, 13
\bibitem[\protect\citeauthoryear{Cappellari et al.}{2006}]{cap06} Cappellari, M., et al. 2006, MNRAS, 366, 1126

\bibitem[\protect\citeauthoryear{Combes et al.}{1990}]{com90} Combes F., Debbasch, F, Friedli, D., Pfenniger, D., 1990, A\&A, 233, 82 
\bibitem[\protect\citeauthoryear{Contopoulos \& Papayannopoulos}{1980}]{con80} Contopoulos, G., Papayannopoulos, T. 1980, A\&A, 92, 33
\bibitem[\protect\citeauthoryear{Copin}{2002}]{cop02} Copin, Y. 2002, in Galaxies: The Third Dimension, ASP Conference Proceedings, Vol. 282, eds Margarita Rosado, Luc Binette, and Lorena Arias
\bibitem[\protect\citeauthoryear{Davies, Sadler \& Peletier}{Davies et al.}{1993}]{dav93} Davies, R. E., Sadler, E. M., Peletier, R. F. 1993, MNRAS, 262, 650
\bibitem[\protect\citeauthoryear{de Zeeuw et al.}{2002}]{dezeeuw02} de Zeeuw, P. T., et al. 2002, MNRAS, 329, 513
\bibitem[\protect\citeauthoryear{Elmegreen}{1994}]{elm94} Elmegreen, B. G. 1994, ApJ, 425, L73
\bibitem[\protect\citeauthoryear{Emsellem et al.}{2004}]{em04} Emsellem, E., et al. 2004, MNRAS, 352, 721
\bibitem[\protect\citeauthoryear{Eskridge et al.}{2000}]{esk00} Eskridge, P. B., et al. 2000, AJ, 119, 536
\bibitem[\protect\citeauthoryear{Falc\'{o}n-Barroso}{2002}]{fal02} Falc\'{o}n-Barroso, J. 2002, PhD Thesis, University of Nottingham
\bibitem[\protect\citeauthoryear{Falc\'{o}n-Barroso et al.}{2006}]{fal06} Falc\'{o}n-Barroso, J. et al., 2006, MNRAS, 369, 529 
\bibitem[Fazio et al.(2004)]{2004ApJS..154...10F} Fazio, G.~G., et al.\
2004, ApJS, 154, 10

\bibitem[\protect\citeauthoryear{Ferrarese et al.}{1996}]{Fer96} Ferrarese, L., et al. 1996, ApJ, 464, 568
\bibitem[\protect\citeauthoryear{Franx, van Gorkom \& de Zeeuw}{1994}]{fran94} Franx, M., van Gorkom, J. H., de Zeeuw, P. T. 1994, ApJ, 436, 642
\bibitem[\protect\citeauthoryear{Ganda et al.}{2006}]{gan06} Ganda, K., Falc\'{o}n-Barroso, J., Peletier, R. F., Cappellari, M., Emsellem, E., McDermid, R. M., de Zeeuw, P. T., Carollo, C. M. 2006, MNRAS, 367, 46
\bibitem[\protect\citeauthoryear{Gerhard}{1993}]{ger93} Gerhard, O. E. 1993, MNRAS, 265, 213
\bibitem[\protect\citeauthoryear{Grosb$\o$l, Patsis \& Pompei.}{2004}]{gros04} Grosb$\o$l, P., Patsis, P. A., Pompei, E. 2004, A\&A, 423, 849 
\bibitem[\protect\citeauthoryear{Heller \& Shlosman}{1994}]{hel94} Heller, C. H., Shlosman, I. 1994, ApJ, 424, 84
\bibitem[\protect\citeauthoryear{Heller \& Shlosman}{1996}]{hel96} Heller, C. H., Shlosman, I. 1996, ApJ, 471, 143
\bibitem[\protect\citeauthoryear{Hernandez et al.}{2005}]{hern05} Hernandez, O., Carignan, C., Amram, P., Chemin, L., Daigle, O. 2005, MNRAS, 360, 1201
\bibitem[Jogee et al.(2002a)]{2002ApJ...570L..55J} Jogee, S., Knapen, J.~H., 
Laine, S., Shlosman, I., Scoville, N.~Z., Englmaier, P.\ 2002a, ApJL, 
570, L55 
\bibitem[Jogee et al.(2002b)]{2002ApJ...575..156J} Jogee, S., Shlosman, I., 
Laine, S., Englmaier, P., Knapen, J.~H., Scoville, N., Wilson, C.~D.\ 
2002b, ApJ, 575, 156 
\bibitem[\protect\citeauthoryear{Jogee et al.}{2004}]{jogee2004} Jogee, S., et al. 2004, ApJL, 615, L105
\bibitem[Kennicutt et al.(2003)]{2003PASP..115..928K} Kennicutt, R.~C., et
al.\ 2003, PASP, 115, 928
\bibitem[\protect\citeauthoryear{Kennicutt et al.}{2005}]{ken05} Kennicutt, R. C., Lee, J. C., Funes, J. G., SJ, Sakai, S., Akiyama, S. 2005, in Starbursts: From 30 Doradus to Lyman Break Galaxies, ed. R. de Grijs \& R. M. González Delgado (ASSL Vol. 329; Dordrecht: Springer), 187
\bibitem[\protect\citeauthoryear{Knapen et al.}{1993}]{k93} Knapen, J. H., Cepa, J., Beckman, J. E., Soledad del Rio, M., Pedlar, A. 1993, ApJ, 416, 563
\bibitem[\protect\citeauthoryear{Knapen et al.}{1995a}]{k95a} Knapen, J. H., Beckman, J. E., Shlosman, I., Peletier, R. F., Heller, C. H., de Jong, R. S. 1995a, ApJ, 443, L73 (K95a)
\bibitem[\protect\citeauthoryear{Knapen et al.}{1995b}]{k95b} Knapen, J. H., Beckman, J. E., Heller, C. H., Shlosman, I., de Jong, R. S. 1995b, ApJ, 454, 623 (K95b)
\bibitem[\protect\citeauthoryear{Knapen et al.}{2000}]{k00b} Knapen, J. H., Shlosman, I., Heller, R. J., Beckman, J. E., Rozas, M. 2000b, ApJ, 528, 219 (K00)
\bibitem[\protect\citeauthoryear{Knapen et al.}{2000}]{k00a} Knapen, J. H., Shlosman, I., Peletier, R. F. 2000a, ApJ, 529, 93
\bibitem[\protect\citeauthoryear{Knapen}{2005}]{k05} Knapen, J. H. 2005, A\&A, 429, 141
\bibitem[\protect\citeauthoryear{Kormendy \& Kennicutt}{2004}]{kor04} Kormendy, J., Kennicutt, R. C. 2004, ARAA, 42, 603
\bibitem[\protect\citeauthoryear{Krajnovi\'c et al.}{2006}]{kraj06} Krajnovi\'c, D., Cappellari, M., de Zeeuw, P.T., Copin, Y. 2006, MNRAS, 366, 787
\bibitem[\protect\citeauthoryear{Kuntschner et al.}{2006}]{kun06} Kuntschner, H., et al. 2006, MNRAS, 369, 497

\bibitem[\protect\citeauthoryear{Laurikainen \& Salo}{2002}]{lau02} Laurikainen, E., Salo, H. 2002, MNRAS, 337, 1118
\bibitem[\protect\citeauthoryear{Leitherer et al.}{1999}]{leith99} Leitherer, C., et al. 1999, ApJ, 123, 3
\bibitem[\protect\citeauthoryear{McDermid et al.}{2006}]{mcderm06} McDermid, R. M., et al. 2006, NewAR, 49, 521

\bibitem[\protect\citeauthoryear{Maciejewski}{2004a}]{mac2004a} Maciejewski, W. 2004a, MNRAS, 354, 883
\bibitem[\protect\citeauthoryear{Maciejewski}{2004b}]{mac2004b} Maciejewski, W. 2004b, MNRAS, 354, 892
\bibitem[\protect\citeauthoryear{Martinez-Valpuesta, Shlosman \& Heller }{2006}]{mart06} Martinez-Valpuesta, I., Shlosman, I., Heller, C. 2006, ApJ, 637, 214
\bibitem[\protect\citeauthoryear{Mazzuca et al.}{2006}]{maz06} Mazzuca, M. L., Knapen, J. H., Veilleux, S., Regan, M. W. 2006, ApJ, submitted
\bibitem[\protect\citeauthoryear{Pahre et al.}{2004}]{par04} Pahre, M. A., Ashby, M. L. N., Fazio, G. G., Willner, S. P. 2004, ApJS, 154, 229
\bibitem[\protect\citeauthoryear{Peletier}{1989}]{pel89} Peletier, R. F. 1989, PhD Thesis, University of Groningen

\bibitem[\protect\citeauthoryear{Prendergast}{1983}]{pren83} Prendergast, K. H. 1983, in Internal Kinematics and Dynamics of Galaxies, ed. E. Athanassoula, 215
\bibitem[\protect\citeauthoryear{Raha et al.}{1991}]{raha91} Raha, N., Sellwood, J. A., James, R. A., Kahr, F. D. 1991, Nature, 411, 42

\bibitem[\protect\citeauthoryear{Regan, Vogel \& Teuben}{1997}]{reg97} Regan, M. W., Vogel, S. N., Teuben, P. J. 1997, ApJ, 482, L143
\bibitem[Rieke et al.(2004)]{2004ApJS..154...25R} Rieke, G.~H., et al.\
2004, ApJS, 154, 25
\bibitem[Ryder \& Knapen(1999)]{1999MNRAS.302L...7R} Ryder, S.~D., Knapen, J.~H.\ 1999, MNRAS, 302, L7
\bibitem[\protect\citeauthoryear{Ryder, Knapen \& Takamiya}{Ryder et al.}{2001}]{ryd01} Ryder, S. D., Knapen, J. H., Takamiya, M. 2001, MNRAS, 323, 663
\bibitem[\protect\citeauthoryear{Sarzi et al.}{2006}]{sar06} Sarzi, M. et al. 2006, MNRAS, 366, 1151
\bibitem[\protect\citeauthoryear{Schoemakers, Franx \& de Zeeuw}{1997}]{scho97} Schoemakers, R. H. M., Franx, M., de Zeeuw, P. T. 1997, MNRAS, 292, 349
\bibitem[\protect\citeauthoryear{Shlosman, Begelman, Frank}{1990}]{shlos90} Shlosman, I., Begelman, M. C., Frank, J. 1990, Nature, 345, 679
\bibitem[\protect\citeauthoryear{Terlevich et al.}{2004}]{terl04} Terlevich, R., Silich, S., Rosa-Gonz\'alez, D., Terlevich, E. 2004, MNRAS, 348, 1191
\bibitem[\protect\citeauthoryear{Thomas, Maraston \& Bender}{Thomas et al.}{2003}]{thom03} Thomas, D., Maraston, C., Bender, R. 2003, MNRAS, 339, 897
\bibitem[\protect\citeauthoryear{Thomas, Maraston \& Korn}{Thomas et al.}{2004}]{thom04} Thomas, D., Maraston, C., Korn, A. 2004, MNRAS, 351, L19 (TM04)
\bibitem[\protect\citeauthoryear{Thomas \& Davies}{2006}]{thom06} Thomas, D. \& Davies, R. L. 2006, MNRAS, 266, 510
\bibitem[\protect\citeauthoryear{van der Marel \& Franx}{1993}]{van93} van der Marel, R. P., Franx, M. 1993, 407, 525
\bibitem[\protect\citeauthoryear{Vazdekis}{1999}]{vaz99} Vazdekis, A. 1999, ApJ, 512, 224
\bibitem[\protect\citeauthoryear{Veilleux \& Osterbrock}{1987}]{vei87} Veilleux, S., Osterbrock, D. E. 1987, ApJS, 63, 295
\bibitem[Werner et al.(2004)]{2004ApJS..154....1W} Werner, M.~W., et al.\
2004, ApJS, 154, 1
\bibitem[\protect\citeauthoryear{Wong, Blitz \& Bosma}{Wong et al.}{2004}]{wong02} Wong, T., Blitz, L., Bosma, A. 2004, ApJ, 605, 183
\bibitem[\protect\citeauthoryear{Worthey, Faber \& Gonzalez}{Worthey et al.}{1992}]{wor92} Worthey, G., Faber, S. M., Gonzalez, J. J. 1992, ApJ, 398, 69
\bibitem[\protect\citeauthoryear{Worthey}{1994}]{wor94} Worthey, G. 1994, ApJS, 95, 107
\end{thebibliography}
\end{document}